%% file: main.tex
\documentclass{article}

\usepackage{arxiv}

\usepackage[utf8]{inputenc}
\usepackage[T1]{fontenc}    
\usepackage{hyperref}       
\usepackage{url}            
\usepackage{booktabs}       
\usepackage{amsfonts}       
\usepackage{nicefrac}       
\usepackage{microtype}      
\usepackage{lipsum}		
\usepackage{graphicx}
\usepackage{natbib}
\usepackage{doi}
\usepackage{framed,multirow}

\usepackage{amssymb}
\usepackage{latexsym}
\usepackage{amsmath}
\usepackage{amsthm}
\usepackage{mathtools}
\usepackage{algorithm2e}
\usepackage{float}
\usepackage{transparent}
\usepackage{makecell}
\usepackage{array}
\usepackage{ulem}
\usepackage{caption}
\usepackage{subcaption}
\usepackage{tikz}

\usepackage{url}
\usepackage{xcolor}
\usepackage{soul}
\usepackage{natbib}
\usepackage{arydshln}

\usepackage{hyperref}
\definecolor{newcolor}{rgb}{.8,.349,.1}

\setstcolor{blue}

\newcommand{\qref}[1]{Eq.~(\ref{#1})}

\newcommand{\E}[1]
{\mathbb{E}\left[#1\right]}

\newcommand{\bE}{\operatorname{\mathbb{E}}}

\newcommand{\wpp}{\Tilde{w}}

\newcommand{\elbo}{\operatorname{ELBO}}

\newtheorem{remark}{Remark}

\title{An information field theory approach to Bayesian state and parameter estimation in dynamical systems}

\date{June 3, 2023}	

\author{{\includegraphics[scale=0.06]{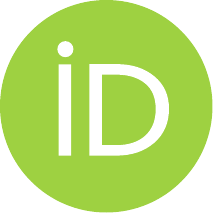}\hspace{1mm}Kairui~Hao} \\
	School of Mechanical Engineering\\
	Purdue University\\
	West Lafayette, IN \\
	\texttt{hao55@purdue.edu} \\
	\And
	{\includegraphics[scale=0.06]{orcid.pdf}\hspace{1mm}Ilias~Bilionis}\thanks{Corresponding author.} \\ 
	School of Mechanical Engineering\\
	Purdue University\\
	West Lafayette, IN \\
	\texttt{ibilion@purdue.edu} \\
}

\renewcommand{\shorttitle}{}

\hypersetup{
pdftitle={A template for the arxiv style},
pdfsubject={q-bio.NC, q-bio.QM},
pdfauthor={David S.~Hippocampus, Elias D.~Striatum},
pdfkeywords={First keyword, Second keyword, More},
}

\begin{document}
\maketitle

\begin{abstract}
	Dynamical system state estimation and parameter calibration problems are ubiquitous across science and engineering.
Bayesian approaches to the problem are the gold standard as they allow for the quantification of uncertainties and enable the seamless fusion of different experimental modalities.
When the dynamics are discrete and stochastic, one may employ powerful techniques such as Kalman, particle, or variational filters.
Practitioners commonly apply these methods to continuous-time, deterministic dynamical systems after discretizing the dynamics and introducing fictitious transition probabilities.
However, approaches based on time-discretization suffer from the curse of dimensionality since the number of random variables grows linearly with the number of time-steps.
Furthermore, the introduction of fictitious transition probabilities is an unsatisfactory solution because it increases the number of model parameters and may lead to inference bias.
To address these drawbacks, the objective of this paper is to develop a scalable Bayesian approach to state and parameter estimation suitable for continuous-time, deterministic dynamical systems.
Our methodology builds upon information field theory.
Specifically, we parameterize the dynamical system state path using a linear basis.
Then, we construct a physics-informed prior probability measure for the coefficients of the linear basis such that functions that satisfy the physics are more likely.
This prior allows us to quantify model form errors.
We connect the system's response to observations through a probabilistic model of the measurement process.
The joint posterior over the system responses and all parameters is given by Bayes' rule.
To approximate the intractable posterior, we develop a stochastic variational inference algorithm.
In summary, the developed methodology offers a powerful framework for Bayesian estimation in dynamical systems.
\end{abstract}


\input{section-introduction}

\input{section-methodology}

\input{section-numerical_examples}
\input{section-conclusions}

\section*{Acknowledgments}
This work was supported by a Space Technology Research Institutes Grant (number 80NSSC19K1076) from NASA’s Space Technology Research Grants Program.

\input{section-appendix}
\bibliographystyle{unsrtnat}

\end{document}

%% file: section-introduction.tex
\section{Introduction}

The goal of dynamical system state estimation is to reveal the time evolution of the state of a dynamical system using a finite amount of noisy measurement data.
More often than not, the underlying physics includes unknown parameters that one needs to identify using the same data.
Such problems are ubiquitous in science and engineering with applications as diverse as
fault detection and diagnostics~\citep{koller2001sampling}, financial engineering~\citep{lautier2003filtering},
computer vision~\citep{szeliski2010computer},
computational biology~\citep{lillacci2010parameter}
,
robotics~\citep{thrun2002probabilistic},
chemical engineering~\citep{oisiovici2000state},
and building air conditioning systems~\citep{bonvini2014robust, hao2020comparing, hao2022comparing}.

The Kalman filter (KF) estimates the state of linear dynamical systems that exhibit Gaussian process noise using linear observations with additive Gaussian noise~\citep{kalman1960new}.
The posterior over states is analytically tractable and one can compute it efficiently by processing the data sequentially in a forward (filter) and backward (smoother) pass~\citep{byron2004derivation}.
The extended Kalman filter (EKF) \cite{anderson2012optimal} deals with non-linear dynamics, albeit with additive Gaussian noise, through sequential local linearizations.
The unscented Kalman filter (UKF) constructs more accurate predictive statistics than the EKF by evaluating the transition function at multiple deterministic points around the current state mean, the so-called sigma points.
The ensemble Kalman filter \citep{evensen2003ensemble} combines Monte Carlo sampling with Kalman filter update rules and can result in improvements over UKF.
Gaussian sum filters are able to approximate non-Gaussian predictive and posterior distributions by running multiple EKFs in parallel and aggregating the posterior estimation information~\cite{anderson2012optimal,wan2000unscented}.
The major drawback of methodologies based on variations of the KF is that they rely on the conjugacy of Gaussian prior and posterior distributions. 
Specifically, for EKF, when the state dynamics and measurement functions exhibit strong nonlinearities,  the Gaussian approximation of the predictive and posterior distributions may lead to progressive deviation of the approximate posterior from the true one~\citep{perea2007nonlinearity}.

\citet{liu1998sequential} developed sequential Monte Carlo (SMC), also known as the particle filter (PF),  to solve the filtering problem in generic dynamical systems.
The method applies sequential importance sampling to approximate continuous probability measures with point measures represented by a finite number of weighted samples.
Such particle approximations converge in distribution due to the law of large numbers and, therefore, are applicable to general non-linear and non-Gaussian cases.
However, it is well known that one of the key challenges while designing the PF is constructing an appropriate transition proposal. 
To avoid particle degeneracy, the transition proposal should lead to a predictive distribution that is close to the posterior.
\citet{pitt1999filtering} modified the standard PF with auxiliary variables to improve tail behaviors in the predictive distribution.
\citet{bocquet2010beyond} and \citet{doucet2001sequential} discussed the optimal proposal that should minimize the variance of particle weights, which is however intractable to implement.
\citet{daum2010exact} proposed an approach that moves the particles from the predictive distribution to the posterior through flows described by ordinary differential equations, thus replacing the steps of using likelihoods to weigh possibly degenerate predictive particles. 
PFs are easily parallelizable so they are a viable choice if one has adequate computational resources available~\citep{goodrum2010parallelization}.

Variational inference (VI) is a type of approximate Bayesian inference that can be used to create filters that are between KF and PF in terms of approximation capabilities and computational cost.
The idea underlying VI is to approximate a probability measure within a parameterized class of probability measures, also known as guides.
VI constructs the approximation by minimizing the Kullback-Liebler (KL) divergence \cite{kullback1951information} between the class of guides and the desired probability measure.
Practitioners can choose the class of guides to balance approximation errors and computational costs.
The simplest choice, the mean field approximation \cite{opper2001advanced}, uses a multivariate Gaussian with diagonal covariance matrix as the guide.
Between the mean field approximation and the full covariance multivariate Gaussian, one can construct guides with a $k$-rank covariance matrix~\citep{seeger2010gaussian}.
Structured mean field \cite{saul1995exploiting} explores appropriate dependence structures of state variables to enlarge the class of guides.
\citet{lund2021variational} explored VI in the context of nonlinear system state and parameter estimation using a multivariate Gaussian distribution with a parameterized banded covariance matrix as the guide.
\citet{archambeau2007variational} considered using a Gaussian process as the guide to infer the posterior distribution of continuous paths in diffusion processes.
\citet{ryder2018black} applied the idea of black box variational inference approach \citep{ranganath2014black} to 
learn the parameters and diffusion paths for
stochastic differential equations (SDEs).
They worked on a discretized version of SDEs through the Euler-Maruyama scheme and chose recurrent neural networks as their guides.

The application of KF, PF or VI to dynamical system state and parameter estimation relies on the time-discretization of the underlying dynamical system.
Time-discretization may introduce inference biases.
Gradient matching, which attempts to perform inference by minimizing the loss between the state derivatives and vector fields, avoids time-discretization and can reduce such biases.
\citet{gorbach2017scalable} exploited gradient matching in the context of VI in a problem where the vector field of a dynamical system was modeled by polynomial functions. They used two exponential families as guides for dynamical system parameters and states, respectively.

Physics-informed machine learning, an emerging field that attempts to combine known physics with state-of-the-art machine learning, can also be thought as a version of gradient matching.
The underlying idea is to construct physics-informed regularizers from differential equations, e.g., using the integrated squared residual of a differential equation~\citep{lagaris1998artificial}.
The rapid development of deep learning software, e.g., TensorFlow~\cite{tensorflow2015-whitepaper}, PyTorch~\cite{NEURIPS2019_9015}, and JAX~\cite{jax2018github}, and hardware, e.g., GPUs~\cite{choquette2021nvidia}, has renewed interest in such approaches and especially on their applications to inverse problems. 
Most common are deterministic approaches that parameterize physical fields with neural networks and train their parameters by minimizing the sum of a data-driven loss and a physics-informed regularizer.
For example, \citet{raissi2019physics,raissi2020hidden} developed physics-informed neural networks (PINNs) to solve inverse problems involving nonlinear partial differential equations.
\citet{chen2020physics} applied PINNs to inverse scattering problems in photonic metamaterials and nano-optics technologies.
\citet{hennigh2021nvidia} solved an inverse problem of a three-fin heat sink, where the task was to estimate flow characteristics such as flow viscosity.
\citet{shukla2020physics} applied PINNs to identify and characterize a surface breaking crack in a plate.

The Bayesian formulation of inverse problems is the gold standard since it can quantify uncertainty and fuse data from heterogeneous information sources. 
To this end, \citet{yang2021b} developed Bayesian PINNs (B-PINNs).
The idea behind B-PINNs is the introduction of fictitious observations that enforce the physics.
Specifically, they assume that the squared residuals at a set of collocation points are an observed quantity with additive Gaussian noise (the observed value is zero).
The variance of this fictitious measurement controls one's trust in the physical equations being correct.
Subsequently, one selects priors for the network weights (typically a zero mean Gaussian) and the physical parameters and forms the joint posterior.
Then, one characterizes this joint posterior either by sampling or VI.
\citet{meng2021multi} expanded this idea by introducing multi-fidelity Bayesian neural networks, which combine PINNs and Bayesian PINNs. They learned the deterministic deep neural networks with low-fidelity data, and used B-PINNs to learn the correlations between the low and high-fidelity data. 
B-PINNs depend on parameterizing the underlying physical fields with neural networks and their limiting behavior is not known.
The use of fictitious observations in B-PINNS to enforce the physics raises some issues.
First, one would expect physical information to appear when specifying prior knowledge -- not as part of the measurement model.
Second, Bayesian asymptotic theory, see theorem on p. 589 of~\cite{gelman2013bayesian}, dictates that the posterior over fields collapses to a point when the number of collocation points goes to infinity.
This behavior is desirable when one has absolute confidence in the model, but not when there is model-form uncertainty.

Information field theory (IFT) \citep{ensslin2009information, ensslin2013information} employs statistical field theory ideas to equip the space of physical fields, e.g., temperature, pressure, velocity fields, with a probability measure which encodes certain beliefs about their characteristics, e.g., smoothness, length-scale. 
This is a probability measure over a function space which is given meaning through the path integrals of~\citet{feynman2010quantum}.
The prior probability measure over the fields is conditioned on data via a probabilistic model of the measurement process.
IFT has been applied successfully to cosmological field inference 
(\cite{tsouros2023reconstructing, platz2022multi, welling2021reconstructing, arras2018radio, porqueres2017cosmic}).
Recent progress has also been made in applying IFT to dynamical systems \cite{frank2021field,westerkamp2021dynamical}.

\citet{alberts2023physics} constructed the prior measure over fields using the fact that they should satisfy given physical laws encoded in differential equations.
Their theory includes a parameter that controls one's prior beliefs about the correctness of the model, a step towards quantifying model-form uncertainty in a principled manner.
If this parameter is zero, then one has no trust in the physics and the method reverts to traditional IFT.
As this parameter increases to infinity, the field prior collapses to the solution of the underlying differential equation.
The authors also demonstrated how IFT can be used in the context of inverse problems, including the problem of estimating the parameter that controls model-form uncertainty.
In addition, they demonstrated that traditional PINNS as well as B-PINNS are limiting cases of IFT.
Despite the theoretical progress in this field, there remain several gaps.
First, the authors restricted their attention only to boundary value problems for which there exists a known energy principle.
In particular, application to dynamical systems is not trivial.
Second, they relied on a nested version of stochastic gradient Langevin dynamics (SGLD)~\citep{welling2011bayesian} to sample from the joint field and parameter posterior.
This sampling scheme is slow-mixing and, thus, computationally demanding.

In summary, the following knowledge gaps remain open in the problem of dynamical system state and parameter estimation.
Despite the continuous nature of ordinary differential equations, traditional and present Bayesian state and parameter estimation methods, including Kalman, particle and VI filters, rely on time-discretization which may introduce unexpected biases.
Next, even though B-PINNs work over continuous function spaces, the introduction of fictitious observations is not desirable in the presence of model-form uncertainty.
Finally, characterizing the IFT posterior via sampling is computationally demanding and inhibits the methods applicability in online settings.

The objective of the paper is to extend IFT in the context of dynamical systems in a computationally efficient manner.
The unique contributions of this work are as follows.
First, we develop a finite-dimensional IFT by parameterizing the infinite-dimensional dynamical system state path using the span of a linear basis.
We introduce the physics-informed conditional prior for the state path parameterization.
The construction of this prior is inspired by the path integral \citep{feynman2010quantum}, and resembles energy-based models \citep{grenander1994representations}.
This prior contains a tunable parameter that plays the role of an inverse temperature, and we demonstrate numerically that it quantifies model form errors.
We connect the proposed physics-informed prior to the joint probability density of dynamical system states of a certain Euler-Maruyama discretized stochastic differential equation. 
Second, we derive a VI scheme to approximate the joint state and parameter posterior within a given set of parameterized distributions.
Note that traditional VI is not directly applicable to our problem because the physics-informed prior includes an intractable normalization constant which depends on the physical parameters.
To circumvent this situation,
we develop a stochastic variational approximation that incorporates a non-convergent and persistent inner loop targeting the physics-informed prior.
The developed algorithm can readily scale to large datasets.
Lastly, we conducted several numerical experiments to compare our method with Bayesian MCMC, particle filters, and B-PINNs.
The numerical examples showcase IFT's ability to accelerate computation and quantify model form errors.

The structure of the paper is as follows.
In section 2, we provide a review of the basics of information field theory.
In section 3,  we develop our theoretical and numerical methodology.
In section 4,  we demonstrate our methodology through numerical examples.
Finally, we conclude the paper in section 5.

%% file: section-methodology.tex
\section{Methodology}
\newcommand{\la}{\Big\langle}
\newcommand{\ra}{\Big\rangle}
\newcommand{\bv}{\big\vert}
\newcommand{\tp}{t^{\prime}}
\label{sec:methodology}

We consider the dynamical system governed by the ordinary differential equation
\begin{align}
    \dot{x}(t) &= f(x(t), t;\theta),\label{ode}
    \\
    Y(t_k) &= R(x(t_k); \theta) + \text{noise},\nonumber
\end{align}
where 
$x(t)\in \mathbb{R}^{d_x}$ is the state variable, 
$f$ is the vector field,
$\theta\in \mathbb{R}^{d_{\theta}}$ is the model parameters, 
$R$ is the measurement response function and
$Y(t_k)\in \mathbb{R}^{d_y}$ is the random output variable.
The noise is typically independent, identically distributed, zero-mean Gaussian.
However, non-Gaussian noise terms can be accommodated if needed.

In the framework of IFT, we parameterize the state function 
$x(t)=\hat{x}(t; w, x_0)$ using a finite number of parameters $w\in \mathbb{R}^{d_w}$ and the initial state $x_0$, i.e., $\hat{x}(0; w, x_0) = x_0$.
This parameterization automatically enforces the initial condition, i.e.,
$\hat{x}(0; w, x_0)=x_0$.
The objective of IFT is to estimate the parameterized state function $\hat{x}(\cdot; w, x_0)$ and the model parameter $\theta$ given a time series measurement data $y=(y(t_1), \cdots, y(t_{n_d}))$.

IFT is a Bayesian inference method applied to physical fields.
Applying Bayes' rule, IFT aims to find the posterior distribution
\begin{align*}
    p(w, x_0, \theta\vert y)=
    \frac{
    p(y\vert w, x_0, \theta)p(w, x_0, \theta)
    }{
    p(y)
    }.
\end{align*}
The likelihood function $p(y\vert w, x_0, \theta)$ can be constructed with the conditionally independent assumption
\begin{align*}
    p(y\vert w, x_0, \theta)
    =
    \prod_{i=1}^{n_d}
    p(y_i\vert \hat{x}(t_i; w, x_0)),
\end{align*}
and each $p(y_i\vert \hat{x}(t_i;w, x_0))$ can be as simple as a multivariate normal.
The unique feature of IFT is the construction of the prior distribution $p(w, x_0, \theta)$, see next section.


\subsection{
Physics-informed prior
}

\label{sec:pif_measure}

We factorize the prior distribution into
\begin{align*}
    p(w, x_0, \theta) = p_{\beta}(w\vert x_0, \theta)p(x_0)p(\theta),
\end{align*}
where $p_{\beta}(w\vert x_0, \theta)$ is
the physics-informed conditional prior defined by
\begin{align}
\label{conditionalprior}
    p_{\beta}(w\vert x_0, \theta)
    =
    \frac{
    e^{-\beta H(w\vert x_0, \theta)}
    }{
    Z_{\beta}(x_0, \theta)
    }.
\end{align}
In the above definition, 
$H(w\vert x_0, \theta)$ is the prior \textit{information Hamiltonian}.
Ideally, it should be derived from a variational reformulation of the physics~\citep{alberts2023physics}. 
When there is not a variational principle, e.g., in the case of a mechanical system with damping and external control input signals, we use the integral of the squared residuals between the state derivative and the vector field to construct it, i.e.,
\begin{align}
\label{cpih}
    H(w\vert x_0, \theta) 
    = 
    \int_0^T dt\ \lVert
     \dot{\hat{x}}(t;w,  x_0)-f(\hat{x}(t;w, x_0), t;\theta) \lVert^2.
\end{align}
The expression $\lVert v\rVert$ is the Euclidean norm of a vector $v$.
The normalization constant
\begin{align*}
    Z_{\beta}(x_0, \theta) = 
    \int_{\mathbb{R}^{d_w}} dw 
    \
    e^{-\beta H(w\vert x_0, \theta)}
\end{align*}
is called the \textit{partition function}.
It ensures that the physics-informed conditional prior is properly normalized.
The hyper-paraemeter $\beta$ is called \textit{inverse temperature} in statistical field theory.

It is worth noting that the set 
$
\{
\hat{x}(t;w, x_0) |\ H(w\vert x_0, \theta)=0
\}
$
denotes the
solutions of the initial value problem $\dot{x} = f(x, t; \theta)$ with $x(0) = x_0$.
If the initial value problem is unique and the parameterization $\hat{x}(t;w, x_0)$ is rich enough, then this set is a singleton.
Hence, the prior probability measure of~\qref{conditionalprior} assigns higher probability to fields that satisfy the physics.
Also, observe that the hyperparameter $\beta$ controls the uncertainty of this prior probability measure. 
Setting $\beta=0$ makes the prior probability measure uniform.
As $\beta$ goes to infinity, the prior probability measure collapses to the ODEs solution set.
Hence, we can interpret $\beta$ as a parameter that represents our trust in the physical model.

\subsection{
Parameterization of sliced function spaces using the linear span of a finite basis
}

To proceed further, let us make a specific choice for the parameterization $\hat{x}(t;w, x_0)$.
We choose to work with a linear span of a finite basis that starts at the initial condition $x_0$.
\begin{remark}
    The choice of parameterizing $\hat{x}(t;w, x_0)$ using a linear basis allows us to safely write the physics-informed conditional prior  $p_{\beta}(w\vert x_0, \theta)$ in \qref{conditionalprior} without adding a determinant of a Jacobian term.
    Then, we can connect  $p_{\beta}(w\vert x_0, \theta)$ to an Euler-Maruyama discretized stochastic differential equation.
    We provide this connection in \ref{appendix:sde}.
\end{remark}

Let $\Tilde{\psi}(t)$ be the $K+1$ basis functions:
\begin{align*}
   \Tilde{\psi}(t) = \left[
        \psi_0(t), 
        \cdots,
        \psi_{K}(t)
    \right]^T.
\end{align*}
We denote the $d_x\times (K+1)$ basis coefficient matrix associated to $\Tilde{\psi}(t)$ by 
$
\Tilde{W}=[w_0, w_1, \dots, w_{K+1}]
$, where each $w_i$ is a $d_x$-dimensional vector.
\begin{remark}
    The specific choice of $\Tilde{\psi}(t)$ depends on individual problems.
    For example, one can use the radial basis if the time evolution state functions have more local variations.
    Alternatively, if the time evolution state functions exhibit periodic behavior, the Fourier basis is a suitable choice.
\end{remark}
\begin{remark}
    The number $K$ of bases is also problem-dependent.
    When the length of the time interval $[0, T]$ is large or the solution of the dynamical system in \qref{ode} contains high-frequency components, one should use more basis terms to achieve an acceptable accuracy.
\end{remark}

To automatically satisfy the initial condition, we explicitly parameterize the function using its initial point $x_0$ such that
$\hat{x}(0;w, x_0)=x_0$.
This can be achieved by solving the equation:
\begin{align*}
    \sum_{i=0}^K
    w_i \psi_i(0) = x_0.
\end{align*}
Then choose one basis $\psi_i(t)$ ($\psi_i(0)\ne 0$) as the dependent basis to compute its coefficient vector:
\begin{align}
    w_i = \frac{
    x_0-\sum_{j\neq i}w_j\psi_j(0)
    }{\psi_i(0)}.
    \label{wj}
\end{align}

We use the notation $W = [w_0, \cdots, w_{i-1}, w_{i+1},\cdots, w_{K}]$ to denote the remaining free coefficient matrix.
If we vectorize this coefficient matrix into a $d_w=d_x\times K$- dimensional vector $w = \operatorname{Vec}(W)$,
we can concisely denote the function in \qref{wj} by $w_i = \mathcal{T}(x_0; w)$.
So the parameterized function is
\begin{align}
    \hat{x}(t; w, x_0)
    =
   \mathcal{T}(x_0; w)
   \psi_i(t)
    +
    \sum_{j\neq i} w_j\psi_j(t).
    \label{repara_x}
\end{align}
It is easy to check this parameterized function automatically satisfies the initial condition, i.e., $\hat{x}(0; w, x_0) = x_0$.

In this way,  the physics-informed conditional prior is a conditional distribution of the $d_w$-dimensional basis coefficients $w$ given the initial point $x_0$ and the vector field parameters $\theta$. From Eqs.~(\ref{cpih}) and (\ref{repara_x}), we formulate the conditional prior information Hamiltonian:
\begin{align}
    H(w\vert x_0, \theta) = \int_0^T dt \ \left\Vert 
    \mathcal{T}(x_0; w)
   \dot{\psi}_i(t)
    +
    \sum_{j\neq i} w_j\dot{\psi}_j(t)
    -
    f\left(
    \mathcal{T}(x_0; w)
   \psi_i(t)
    +
    \sum_{j\neq i} w_j\psi_j(t)
    , t;\theta
    \right) \right\lVert^2,
    \nonumber
\end{align}
to construct the physics-informed conditional prior 
$
p_{\beta}(w|x_0, \theta)= 
\frac{
e^{-\beta H(w|x_0, \theta)}
}{
Z_\beta(x_0, \theta)
}
$.
Then the joint probability density has the form of
\begin{align}
    p_\beta(y,w, x_0, \theta) 
    =
    \frac{
        p(y\vert w, x_0, \theta)
        e^{
            -\beta H(w|x_0, \theta)
        }
        p(x_0, \theta)
    }
    {Z_\beta(x_0, \theta)},
    \label{jointdistribution}
\end{align}
and the posterior probability measure is
\begin{equation}
p_\beta(w, x_0, \theta \vert y) 
    =
    \frac{
    p_{\beta}(y, w, x_0, \theta) 
    }
    {p(y)}
    \nonumber
    =
    \frac{
        p(y\vert w, x_0, \theta)
        e^{-\beta H(w\vert x_0, \theta)}
        p(x_0,\theta)
    }
    {p(y)Z_\beta(x_0, \theta)}.
    \label{posterior}
\end{equation}
\begin{remark}
    Observe the appearance of the intractable normalization constant $Z_\beta(x_0, \theta)$. 
    The posterior distribution takes greater values when the likelihood $p(y\vert w, x_0, \theta)$, the unnormalized physics-informed conditional prior $e^{-\beta H(w\vert x_0, \theta)}$ and the prior $p(x_0, \theta)$ are greater, plus the partition function $Z_{\beta}(x_0, \theta)$ is smaller.
    This means we want to find $w$, $x_0$, and $\theta$ to agree with the data and the prior while keeping the total field free energy, i.e., the partition function, small.
\end{remark}

\subsection{
The necessity of the partition function $Z_{\beta}(x_0, \theta)$:
Comparison to B-PINNs from a hierarchical Bayesian model perspective
}

We compare our IFT formulation to B-PINNs formulation from a hierarchical Bayesian model perspective.
First, we translate the B-PINNs formulation \citep{yang2021b} into the problem of dynamical system state and parameter estimation.
We parameterize the time evolution states by a function $\hat{x}(t; w, x_0)$. 
In the framework of B-PINNs, one generates the fictitious residual observation data set $y^r = (y^r_1, \cdots, y^r_{n_{r}})$, where each $y^r_{i}$ is a $d_x$-dimensional vector which follows the Normal distribution with mean centered at the ODE residual and variance $\sigma^2_{y^r}$. 
For the ``observed'' $y^r$, one uses all zeros.
The likelihood function for these fictitious residual measurements then has the form
\begin{align}
    p(y^r=0|w, x_0, \theta) 
    \propto
    \exp\left\{
        -\frac{1}{2\sigma_{y_r}^2}\sum_{i=1}^{n_r}
                \left\Vert \dot{\hat{x}}(t_i; w, x_0)-
                f\left(
                \hat{x}(t_i;w, x_0), t_i; \theta
                \right)
                \right\Vert^2
    \right\}. 
    \label{logpyr}
\end{align}
Typically, one picks a zero-mean Gaussian prior for the weights: $p(w)\propto \exp\{-\alpha \Vert w\Vert^2\}$.
So, B-PINNs write the inverse problem as
\begin{align}
    p(w, x_0, \theta| y, y^r=0)
    &\propto
    p(y|w, x_0, \theta)
    p(y^r=0|w, x_0, \theta)
    p(w)p(x_0)p(\theta).
    \label{bpinns_formulation}
\end{align}
As a comparison, IFT formulates an inverse problem that has the form 
\begin{align}
    p(w, x_0, \theta| y) 
    \propto
    p(y|w, x_0, \theta)p_{\beta}(w|x_0, \theta)p(x_0)p(\theta).
\end{align}
So, the difference lies in replacing the fictitious residual observations of B-PINNs, i.e., $p(y^r=0|w,x_0,\theta)p(w)$, with the physics-informed prior of IFT, i.e., $p_\beta(w|x_0,\theta)$:
$$
(\text{B-PINNs})\;\;\;p(y^r=0|w,x_0,\theta)p(w)\longrightarrow p_\beta(w|x_0,\theta)\;\;\;(\text{IFT}).
$$
Closer inspection reveals that the real difference is the partition function.
To see this, assume that the fictitious observation timestep is fixed to $\Delta t$.
Then, we can connect the fictitious B-PINNs likelihood to the IFT information Hamiltonian as follows:
\begin{align}
\log p(y^r=0|w,x_0,\theta)
    &= 
    -\frac{1}{2\sigma_{y_r}^2}\sum_{i=1}^{n_r}\left\Vert \dot{\hat{x}}(t_i; w, x_0)-
                f\left(
                \hat{x}(t_i;w, x_0), t_i; \theta
                \right)
                \right\Vert^2 + \text{const}\\
    &=-\frac{1}{2\sigma_{y_r}^2\Delta t}\int_0^Tdt \left\Vert \dot{\hat{x}}(t; w, x_0)-
                f\left(
                \hat{x}(t_i;w, x_0), t; \theta
                \right)
                \right\Vert^2 + O\left(\Delta t^2\right) + \text{const}\\
    &= -\frac{1}{2\sigma_{y_r}^2\Delta t}H(w|x_0,\theta) + O\left(\Delta t^2\right) + \text{const}.
\end{align}
Setting $\beta = \frac{1}{2\sigma_{y_r}^2\Delta t}$, we formally obtain IFT from B-PINNs by making the change:
$$
(\text{B-PINNs})\;\;\;p(y^r=0|w,x_0,\theta)p(w) \approx e^{-\beta H(w|x_0,\theta)}p(w) \longrightarrow p_\beta(w|x_0,\theta)=\frac{e^{-\beta H(w|x_0,\theta})}{Z_{\beta}(x_0,\theta)}\;\;\;(\text{IFT}).
$$\\
\indent Equipped with this theoretical comparison between B-PINNs and IFT.
To simplify the discussion, assume $p(w)$ is flat.
we can make the following observations:
\begin{enumerate}
\item B-PINNs and IFT give practically the same posterior over $w$ (for a suitable choice of $\sigma_{y_r}^2$ and $\beta$).
\item B-PINNs and IFT yield different posteriors over initial conditions $x_0$ and the model parameters $\theta$, because B-PINNs are missing the $Z_\beta(x_0,\theta)$ normalization constant.
\end{enumerate}
In section~\ref{sec:compare-to-b-pinns} we numerically demonstrate these theoretical observations.
\\

\subsection{Numerical methods}

An analytical characterization of the posterior distribution \qref{posterior} is rarely possible, so we develop a numerical method to sample from the posterior distribution \qref{posterior}.
Some insights can be gained from the literature on probabilistic energy-based models for density estimation.
For example, \citet{grenander1994representations} showed that maximum likelihood estimation requires sampling from the model distribution, which they carried out using 
Langevin dynamics.
Running Langevin dynamics until convergence is time-consuming, so variants such as contrastive divergence (CD) \citep{carreira2005contrastive}, persistent CD \citep{tieleman2008training}, and short-run MCMC \citep{nijkamp2019learning} have been proposed.
However, these approaches are typically used to model the data distribution instead of the latent variable distribution that appears in our case.
\citet{pang2020learning} trained a latent energy-based model for image generation by short-run MCMC.
The method requires two-stage, i.e., nested sampling. The outter part samples parameters and it relies on inner samplers that sample from the prior and the posterior.

\citet{alberts2023physics} performed Bayesian inference with a physics-informed prior using a specially designed nested SGLD scheme. 
Specifically, their approach includes an outter sampler for the parameters that uses information from two inner, non-convergent, persistent SGLD samplers targeting on the unnormalized prior and the posterior, respectively.
The addendum of this approach is twofold.
First, it scales up to large datasets because it subsamples the data.
Second, it uses a sampling average approximation of the Hamiltonian and thus avoids costly quadrature schemes.
The drawback of this nested SGLD approach is that the posterior parameter samples mix very slowly and a lot of iterations are required for convergence.

To remedy these issues, we developed a stochastic VI approach.
Due to the intractable normalization constant, our algorithm also requires the ability to sample from the prior.
Instead of using a sampling scheme, we fit the prior to a parameterized guide by solving a non-convergent, persistent variational problem.
The resulting algorithm is essentially a maxmini optimization problem.

\subsubsection{Variational inference}

The goal is to perform the inference over the induced finite-dimensional function space $F = \cup_{x_0 \in \mathbb{R}^{d_x}} F_{x_0}$.
In variational inference, one starts by defining a collection of candidate probability densities $\mathcal{Q}$.
An element $q$ of $\mathcal{Q}$ is typically referred to as ``the guide.'' 
Then, one selects a $q$ in $\mathcal{Q}$ by minimizing the KL divergence between it and the posterior $p(w,x_0,\theta|y)$:
\begin{align}
    D_{KL}(q|| \ p_\beta(\cdot|y))
    =
    \int_{
    \mathbb{R}^{d_w}\times \mathbb{R}^{d_x} \times\Theta
    }
     dw dx_0 d \theta \
    \log{
        \frac{q(w, x_0, \theta)}{p_\beta(w, x_0, \theta|y)}
    }
    q(w, x_0, \theta).
    \label{KL_guide_posterior}
\end{align}
Directly minimizing the KL divergence is not possible because its evaluation depends on the intractable posterior.
However, one can show that minimizing the KL divergence is equivalent to maximizing the evidence lower bound (ELBO)~\citep{kingma2013auto}:
\begin{align}
    \operatorname{ELBO}[q] 
    =
    \int_{
        \mathbb{R}^{d_w}\times \mathbb{R}^{d_x} \times\Theta
    }
     dw dx_0 d \theta \
    \log{
       \frac{p_\beta(y, w, x_0, \theta)}
       {q(w, x_0, \theta)}
    }
    q(w, x_0, \theta).
    \label{ELBO_deco}
\end{align}
We see that the ELBO only requires evaluating an expectation of the joint density $p_{\beta}(y, w, x_0, \theta)$.
This still includes the intractable partition function, but it is more manageable than the posterior $p_{\beta}(w,x_0,\theta|y)$.

In practice, one does not maximize \qref{ELBO_deco} over an infinite dimensional probability density space.
Instead, they restrict $\mathcal{Q}$ to be the space spanned by a specific finite dimensional parameterization.
In the sections that follow, we pick a specific guide and develop an algorithm that maximizes~\qref{ELBO_deco}.

\subsubsection{Variational approximation to the physics-informed conditional prior $p_{\beta}\left(w\vert \lambda\right)$}

To simplify notations,  we collectively write $\lambda=\left(x_0, \theta\right)$. 
So we will use $p_{\beta}\left(w\vert \lambda\right)$ and $p_\beta\left(y, w, \lambda\right)$ to denote the physics-informed conditional prior and joint distribution, respectively.

We construct a variational inference approximation to the physics-informed prior.
The first obstacle we need to overcome is the time integration required for evaluating the information Hamiltonian.
To this end, define the function:
\begin{align}
    h_\beta(w, t|\lambda) &:=  T\beta\left\Vert 
    \dot{\hat{x}}\left(t; w, x_0\right) - 
    f\left(\hat{x}\left(t; w, x_0\right), t; \theta\right)
    \right\Vert^2, 
    \label{hwt}
\end{align}
and use it to rewrite the information Hamiltonian $H\left(w| \lambda\right)$ as an expectation over time:
\begin{align}
    H(w|\lambda) 
    =
    \beta^{-1}\int_{0}^T dt \ T^{-1} h_\beta (w, t|\lambda)
    =
     \beta^{-1}\bE_{t\sim \mathcal{U}([0, T])}
     \left[h_\beta\left(w, t| \lambda\right)
     \right]. \nonumber
\end{align}
This alternative view of the information Hamiltonian allows us to construct a sampling average approximation. 

Let $q_\phi$ be the guide we use to approximate $p_\beta(w|\lambda)$.
We find $\phi$ by maximizing the ELBO:
\begin{align}
    \operatorname{ELBO}
    \left(\phi\vert \lambda\right)
    \coloneqq
    -
    \bE_{
    w\sim q_{\phi},
    t\sim \mathcal{U}([0, T])
    }
    \left[
        h_\beta
    \left(
    w,t
    |
    \lambda
    \right)
    \right]
    +
    \mathbb{H}\left[q_{\phi}\right],
    \label{vari_prior}
\end{align}
where $
\mathbb{H}\left[q_\phi\right]= -\bE_{w\sim q_\phi}\left[\log\left(q_\phi(w)\right)\right]
$
is the entropy $q_\phi(w)$.
For a Gaussian guide with covariance $\Sigma$ the entropy is analytically available:
\begin{align*}
    \mathbb{H}\left[q_{\phi}\right]
    =
    \frac{1}{2}
    \log\left\vert\Sigma\right\vert
    +
    \frac{D}{2}\left[1+\log\left(2\pi\right)\right].
\end{align*}
Maximizing \qref{vari_prior} is equivalent to minimizing the KL divergence between $q_{\phi}(w)$ and $p_{\beta}(w\vert \lambda)$ (see \ref{appendix_auxiliary_guide_appro_prior}).

We construct a stochastic gradient optimization variant that maximizes~\qref{vari_prior} over the variational parameters $\phi$.
The required ingredient is an unbiased estimator of the gradient of the ELBO which we can obtain by using the reparameterization trick of~\citet{kingma2013auto}.
The approach parameterizes the guide as a deterministic transform of a fixed base distribution, such as the standard normal, such that the reparameterized guide follows the same distribution as $q_{\phi}\left(w\right)$ with the same variational parameter $\phi$.
Let $\epsilon$ follow a fixed base distribution $q\left(\epsilon\right)$ and $g_{\phi}$ be a deterministic transform map parameterized by $\phi$. 
If the random variable $w =g_{\phi}\left(\epsilon\right)$ follows $q_{\phi}\left(w\right)$, 
we can apply change of variables to \qref{vari_prior}
 and get an expression of the $\operatorname{ELBO}\left(\phi\vert \lambda\right)$ in terms of the random variable $\epsilon$:
\begin{equation}
     \operatorname{ELBO}
    \left(\phi\vert \lambda\right)
    \coloneqq
    -
    \mathbb{E}
    \left[
    h_\beta
    \left(
    g_{\phi}\left(
    \epsilon\right), t
    \Big|
   \lambda
    \right)
  \right]
  +
  \mathbb{H}\left[q_{\phi}\right],\label{vari_prior_rep}
\end{equation}
where the expectation is over $\epsilon\sim q$ and $t\sim \mathcal{U}([0, T])$ and, in particular, it does not depend on $\phi$.
So taking the gradient, we have:
\begin{align}
    \nabla_{\phi}  \operatorname{ELBO}
    \left(\phi\vert \lambda\right)
    =
    -\bE
    \left[
    \nabla_{\phi}\
    h_\beta
    \left(
    g_{\phi}\left(
    \epsilon\right), t
    \Big|
   \lambda
    \right)
    \right]
    +
    \nabla_{\phi}
    \mathbb{H}\left[q_{\phi}(w)\right]
    . \nonumber
\end{align}
Let $\epsilon_i$ and $t_{ij}$ be independent samples of $\epsilon$ and $t$ from $q$ and $\mathcal{U}([0,T])$, respectively.
The unbiased estimator of the gradient of the ELBO is:
\begin{align}
    \widehat{
     \nabla_{\phi}  \operatorname{ELBO}
    \left(\phi\vert \lambda\right)
    }
    =
    -\frac{1}{n_\epsilon n_t}
    \sum_{i=1}^{n_\epsilon}
    \sum_{j=1}^{n_t}
       \nabla_{\phi} \
       h_\beta
       \left(
       g_{\phi}
       \left(
       \epsilon_i
       \right), t_{ij}
       \Big \vert
      \lambda
       \right)
    +
    \nabla_{\phi}
    \mathbb{H}\left[q_{\phi}\right]
    . \label{grad_elbo_prior_estimator}
\end{align}

Equipped with the unbiased estimator of the gradient, we can now use any variant of stochastic gradient ascent (as long as the learning rates satisfy the Robins-Monro conditions \cite{robbins1951stochastic}).
We use an Adam variant \cite{kingma2014adam} summarized in algorithm~\ref{alg:vi_prior}.

\RestyleAlgo{ruled} 
\SetKwInput{KwInit}{Initialization}
\SetKwInput{KwInput}{Input}
\SetKwInput{KwReturn}{Return}
\SetKwInput{Kwalg}{VI\_PRIOR}
\SetKwComment{Comment}{$\triangleright$}{}
\begin{algorithm}[!ht]
\caption{
Variational approximation to $p_{\beta}\left(w\vert \lambda\right)$.
}\label{alg:vi_prior}
\Kwalg{
(
\begin{tabbing}
\hspace{1em} \= $\lambda = \left(x_0, \theta\right)$, \hspace{1em} \=
 \# \textit{The initial state and parameter.}
\\
\> $\phi$, \> \# \textit{Initial variational parameters.}
\\
\> niter, \> \# \textit{The number of optimization iterations.}
\\
\> $(n_\epsilon, n_t)$ \> \# \textit{Sample sizes.}
\\
)
\end{tabbing}
}
 \For{$\text{it}=0; \text{it}<\text{niter}; \text{it}=\text{it}+1$}{  
 Sample $\epsilon_i$ independently from $q$\;
 Sample $t_{ij}$ independently from $\mathcal{U}([0,T])$\;
 Compute $\widehat{
     \nabla_{\phi}  \operatorname{ELBO}
    \left(\phi\vert \lambda\right)
    }$ in \qref{grad_elbo_prior_estimator}
    \;
  Update $\phi$ using Adam with gradient estimate $\widehat{\nabla_{\phi}  \operatorname{ELBO}\left(\phi\vert \lambda\right)}$;
} 
 \KwReturn{
 $
 \phi
 $
 \hspace{2em}
 \#
 \textit{Final variational parameters.}
 }
\end{algorithm}

\subsubsection{Variational approximation to the posterior $p_\beta\left(w, \lambda\vert y\right)$}

We are ready to build a variational inference algorithm to approximate the posterior distribution $p_\beta\left(w, \lambda\vert y\right)$.
Algorithm \ref{alg:vi_prior} will be a building block.
For notational convenience, we write the joint distribution of \qref{jointdistribution} in terms of information Hamiltonians by defining:
\begin{align*}
    H(y|w, \lambda) = -\log\left(p\left(y|w, \lambda\right)\right), \
    H(\lambda) = -\log\left( p\left(\lambda\right)\right),
\end{align*}
and 
\begin{align*}
    H_\beta(y, w, \lambda)
    =
    H(y|w, \lambda) + \beta H(w|\lambda) + H(\lambda).
\end{align*}

Let $q_{\phi}(w)$ and $q_{\psi}(\lambda)$, with variational parameters $\phi$ and $\psi$, be the guides of $w$ and $\lambda$, respectively.
Then the ELBO takes the form of:
\begin{equation}
    \elbo\left(\phi, \psi\vert y\right) 
    =
     -\bE_{w\sim q_{\phi},\lambda \sim q_{\psi}}
    \left[
    H_\beta (y, w, \lambda)
    \right]
    +
    \mathbb{H}\left[
    q_{\phi}\right]
    +
    \mathbb{H}\left[q_{\psi}
    \right]
    -\mathbb{E}_{\lambda \sim q_\psi}\left[\log\left(Z_\beta\left(\lambda\right)\right)\right].
    \label{elbo_post_no_t}
\end{equation}
Similar to Algorithm~\ref{alg:vi_prior}, we can avoid doing the time integration by expressing $H_\beta(y,w,\lambda)$ as an expectation over a uniformly distributed time variable.
To this end, define:
\begin{align}
    h_\beta\left(y, w, \lambda, t \right) &:=
     H(y|w, \lambda) + h_\beta(w, t|\lambda) + H(\lambda), \label{h_beta_joint}
\end{align}
where $h_\beta(w, t|\lambda)$ was defined in \qref{hwt}.
Then:
\begin{align}
     H_\beta(y, w, \lambda) &=
     \bE_{t\sim\mathcal{U}([0, T])}
     \left[
     h_\beta \left(y, w, \lambda, t\right)
     \right].\nonumber
\end{align}
Substituting the equation above into \qref{elbo_post_no_t}, we end up with an ELBO expression that is suitable for constructing a sampling average approximation:
\begin{align}
    \elbo\left(\phi, \psi\vert y\right) 
    =
     -\bE_{w\sim q_{\phi},\lambda \sim q_{\psi},t\sim\mathcal{U}([0, T])}
    \left[
    h_\beta(y, w, \lambda, t)
    \right]
    +
    \mathbb{H}\left[
    q_{\phi}\right]
    +
    \mathbb{H}\left[q_{\psi}
    \right]
    -\mathbb{E}_{\lambda \sim q_\psi}\left[\log\left(Z_\beta\left(\lambda\right)\right)\right]
    . \nonumber
\end{align}
We now apply the reparameterization trick.
Let $\epsilon$ and $\eta$ be the base random variables following the distribution $q$ (e.g., a standard normal) for $w$ and $\lambda$, respectively.
Choose two transformation maps $g_{\phi}(\epsilon)$ and $g_{\psi}(\eta)$ with the properties $w = g_{\phi}(\epsilon)\sim q_{\phi}$ and $\lambda=g_{\psi}\left(\eta\right)\sim q_{\psi}$.
Reparameterizing, we get:
\begin{align}
    \elbo\left(\phi, \psi\vert y\right)
    =
    -\bE
    \left[
    h_\beta\left(y, g_{\phi}\left(\epsilon\right), g_{\psi}\left(\eta\right),
    t
    \right)
    \right]
    +
    \mathbb{H}\left[
    q_{\phi}\right]
    +
    \mathbb{H}\left[q_{\psi}\right]
    -\mathbb{E}\left[\log\left(Z_\beta\left(g_{\psi}\left(\eta\right)\right)\right)\right], \nonumber
\end{align}
where the expectation is over $\epsilon\sim q, \eta\sim q$ and $t\sim\mathcal{U}([0, T])$.
The gradient with respect to $\phi$ is:
\begin{equation}
    \label{elbo_grad_phi}
\nabla_\phi\elbo(\phi, \psi\vert y) = -
\bE
    \left[
    \nabla_\phi h_\beta\left(y, g_{\phi}\left(\epsilon\right), g_{\psi}\left(\eta\right),
    t
    \right)
    \right]
    +
    \nabla_\phi\mathbb{H}\left[q_\phi\right].
\end{equation}
The gradient with respect to $\psi$ is more involved as it requires differentiating through the partition function.
In \ref{appendix_grad_ELBO}, we show that:
$$
\nabla_\psi \mathbb{E}\left[\log\left(Z_\beta\left(g_\psi(\eta)\right)\right)\right] = -\mathbb{E}\left[\mathbb{E}\left[\nabla_\psi h_\beta(\tilde{w},t|g_\psi(\eta))\middle|\eta\right]\right],
$$
where the inner expectation of the right hand side is over the physics-informed conditional prior $\tilde{w}\sim p_\beta(\wpp|\lambda=g_\psi(\eta))$ and the time $t\sim\mathcal{U}([0,T])$, while the outer expectation is over $\eta \sim q$.
The required gradient of the ELBO is:
\begin{equation}
    \nabla_\psi \elbo(\phi, \psi\vert y)
    =
    -
    \bE
    \left[
    \nabla_{\psi}\
    h_\beta\left(
    y, g_{\phi}\left(
    \epsilon
    \right),
    g_{\psi}\left(\eta\right),t
    \right)
    \right]
    +
    \nabla_{\psi}
    \mathbb{H}\left[
    q_{\psi}
    \right]
    +\E{\E{\nabla_\psi h(\tilde{w},t\vert g_\psi(\eta))\middle\vert\eta}}
    .
    \label{elbo_grad_psi}
\end{equation}

We are ready to construct unbiased estimators of the two gradients (Eqs.~(\ref{elbo_grad_phi}) and~(\ref{elbo_grad_psi})).
For the gradient in \qref{elbo_grad_phi} things are straightforward.
Let ($\epsilon_i$, $\eta_i$) and $t_{ij}$ be independent samples of $(\epsilon, \eta) \sim q$ and $t\sim \mathcal{U}([0,T])$, respectively.
We define the sampling average:
\begin{equation}
\widehat{\nabla_\phi \elbo(\phi,\psi\vert y)} = 
-\frac{1}{n_{\epsilon \eta} n_t}\sum_{i=1}^{n_{\epsilon\eta}}\sum_{j=1}^{n_t}\nabla_\phi h_\beta (y, g_\phi(\epsilon_i), g_\psi(\eta_{i}), t_{ij}) + \nabla_\phi\mathbb{H}[q_\phi].
\label{elbo_grad_phi_sa}
\end{equation}

For the gradient of \qref{elbo_grad_psi}  we need to sample $\tilde{w}$ from the field prior $p_\beta(\Tilde{w}\vert \lambda=g_\psi(\eta_{i}))$ for each $\eta_{i}$.
We achieve this in two steps.
First, we run Algorithm~\ref{alg:vi_prior} to obtain a variational approximation $q_{\tilde{\phi}_{i}}$ of the desired prior.
We call this the auxiliary variational inference problem.
Second, we sample independently from the variational approximation $n_{\Tilde{w}}$ times to obtain $\tilde{w}_{is}$.
We define the sampling average:
\begin{equation}
    \begin{array}{ccl}
    \widehat{\nabla_\psi \elbo(\phi,\psi\vert y)} &=& 
    -\frac{1}{n_{\epsilon\eta} n_t}\sum_{i=1}^{n_{\epsilon\eta}}\sum_{j=1}^{n_t}\nabla_\psi h_\beta (y, g_\phi(\epsilon_i), g_\psi(\eta_{i}), t_{ij}) + \nabla_\psi\mathbb{H}[q_\psi]\\
    &&
    +
    \frac{1}{n_{\epsilon\eta} n_{\tilde{w}} n_{\Tilde{t}}}\sum_{i=1}^{n_{\epsilon\eta}}\sum_{s=1}^{n_{\tilde{w}}}
    \sum_{k=1}^{n_{\Tilde{t}}}
    \nabla_\psi h_\beta(\tilde{w}_{is}, t_{isk}\vert  g_\psi(\eta_i))
    .
    \end{array}
    \label{elbo_grad_psi_sa}
\end{equation}
This estimator is biased since we do not draw the samples $\tilde{w}_{is}$ from the exact prior but from the variational approximation of the prior.
In \ref{appendix_maxmini_vi}, we show how stochastic gradient ascent on the ELBO with this estimator is equivalent to solving a maxmini variational inference problem.

\RestyleAlgo{ruled} 
\SetKwInput{KwInit}{Initialization}
\SetKwInput{KwInput}{Input}
\SetKwInput{KwReturn}{Return}
\SetKwInput{Kwalg}{VI\_POSTERIOR}
\SetKwComment{Comment}{$\triangleright$}{}
\begin{algorithm}[!bht]
\caption{
Variational approximation to $p_\beta\left(w, \lambda\vert y\right)$.
}\label{alg:vi_post}
\Kwalg{
(
\begin{tabbing}
\hspace{1em} \= $\left(\phi, \psi, \Tilde{\phi}\right)$, \hspace{4em} \=
 \# \textit{Variational parameters.}
\\
\> (niter, niter\_auxi), \> \# \textit{The number of optimization iterations.}
\\
\> $(n_{\epsilon \eta},n_t, n_{\tilde{w}}, \tilde{n}_\epsilon, \tilde{n}_t)$, \> \# \textit{Sample sizes.}
\\
)
\end{tabbing}
}
 \For{$\text{it}=0; \text{it}<\text{niter}; \text{it}=\text{it}+1$}{
 \# Assuming $n_{\epsilon\eta} = 1$ for computational efficiency\\
 Sample $\eta_1 \sim q$\;
 Compute $\lambda_1 = g_{\psi}\left(\eta_1\right)$
\;
\# Run \textbf{Algorithm \ref{alg:vi_prior}\\
$\Tilde{\phi}\gets$ VI\_PRIOR}
$
\left(
\lambda_1, 
\tilde{\phi},
\text{niter\_auxi},
\left(\tilde{n}_\epsilon, \tilde{n}_t\right)
\right)
$\;
Sample $\epsilon_1$  from $q$\;
Sample $\tilde{w}_{1s}$ independently from $q_{\tilde{\phi}}$\;
Sample $t_{1j}$ and $t_{1sk}$ independently from $\mathcal{U}([0,1])$\;
Compute $\widehat{\nabla_\phi \elbo(\phi,\psi\vert y)}$ using \qref{elbo_grad_phi_sa}\;
Compute $\widehat{\nabla_\psi \elbo(\phi,\psi\vert y)}$ using \qref{elbo_grad_psi_sa}\;
Update $\phi$ using Adam with gradient estimate $\widehat{\nabla_\phi \elbo(\phi,\psi\vert y)}$\;
Update $\psi$ using Adam with gradient estimate $\widehat{\nabla_\psi \elbo(\phi,\psi\vert y)}$\;
 }
 \KwReturn{
$\left(\phi, \psi\right)$
 \hspace{2em}
 \#
 \textit{Approximate posterior distributions.}
 }
\end{algorithm}

Algorithm~\ref{alg:vi_post} summarizes the steps, and we leave some remarks to the readers.
First, running the internal Algorithm~\ref{alg:vi_prior} to convergence is computationally impractical.
Instead, we propose to do a single internal iteration, i.e., $\text{niter\_auxi}=1$, for each outer stochastic gradient ascent iteration.
In our numerical examples, we have found that this choice is sufficient for convergence of the outer scheme and it is much more computationally efficient.
Second, notice that the cost of each outer iteration grows linearly with increasing number of observations in $y$.
In case of independent measurements, one can use common subsampling tricks to obtain an algorithm that scales to large datasets.
See section 4.3.1 in \citet{alberts2023physics} for more details on how this can be achieved.

\subsubsection{The selection of the guide}

The guide balances expressiveness and computational costs.
Above, we have implicitly chosen to work with a guide that factorizes over field parameters $w$ and the tuple of initial conditions and physical parameters $\lambda$, i.e.,
$$
q(w,\lambda) = q_\phi(w)q_\psi(\lambda).
$$
We further factorize over $q_\psi(\lambda)$ as:
$$
q_{\psi}(\lambda) = q_\psi(x_0, \theta) = q_{\psi}(x_0)q_{\psi}(\theta).
$$
We pick the sub-guide $q_{\phi}(w)$ to be a $d_w $-dimensional multivariate normal distribution parameterized by a mean vector $m_w$ and
a diagonal covariance matrix with the diagonal elements
$\sigma^2_{w_{i}}$, i.e.:
\begin{align}
q_{\phi}(w) 
=
\mathcal{N}\left(w\middle|
m_w, \text{diag}\left(\sigma^2_{w_1},\dots,\sigma^2_{w_{d_w}}\right)\right)
\nonumber
\end{align}
The auxiliary guide $q_{\tilde{\phi}}\left(\Tilde{w}\right)$ has the same structure as $q_{\phi}\left(w\right)$:
\begin{align}
q_{\Tilde{\phi}}(\wpp) 
=
\mathcal{N}\left(\wpp\middle|
m_{\wpp}, \text{diag}\left(\sigma^2_{\wpp_1},\dots,\sigma^2_{\wpp_{d_{\wpp}}}\right)\right).
\nonumber
\end{align}

Similarly, for $q_{\psi}(x_0)$ we use the $d_x$-dimensional  multivariate normal distribution parameterized by a mean vector $m_{x_0}$ and a diagonal covariance matrix with the diagonal elements 
$
\sigma^2_{x_{0, i}}
$:
\begin{align}
q_{\psi}(x_0) 
=
\mathcal{N}\left(x_0\middle|
m_{x_0}, \text{diag}\left(\sigma^2_{x_{0, 1}},\dots,\sigma^2_{x_{0, d_x}}\right)
\right).
\nonumber
\end{align}
These simplifying assumptions reduce the computational cost, but they miss correlations between the variables.

We would like to retain information about correlations for the physical parameters $\theta$.
Therefore, we pick $q_{\psi}(\theta)$ to be the full-rank multivariate normal distribution parameterized by the mean vector $m_{\theta}$ and the covariance matrix $\Sigma_{\theta}$:
\begin{align*}
     q_{\psi}(\theta) = \mathcal{N}\left(\theta\middle|m_{\theta}, \Sigma_\theta\right).
\end{align*}

The variational distributions $q_{\phi}(w)$, $q_{\psi}(x_0)$, $q_{\psi}(\theta)$ and $q_{\tilde{\phi}}\left(\Tilde{w}\right)$ have corresponding variational parameters to be optimized. 
Some of them have support constraints, e.g., the variances 
$\{\sigma^2_{w_{i}}\}$, $\{\sigma^2_{x_{0, i}}\}$ and $\{\sigma^2_{\Tilde{w}_{i}}\}$ should be positive,
 and the covariance matrix $\Sigma_{\theta}$ should be positive definite.
We remove these constraints by reparameterizing.
Let $\zeta$ denote any of the $w_i$,  $x_0$ or $\Tilde{w}_i$.
Let $\bar{\sigma}_\zeta$ be the unconstrained version of $\sigma_\zeta$.
We set:
$$
\sigma_{\zeta} = \operatorname{softplus}\left(\bar{\sigma}_\zeta\right) = \log\left(1+e^{\bar{\sigma}_\zeta}\right).
$$
We use $\bar{\sigma}_{w}$,  $\bar{\sigma}_{x_0}$ and $\bar{\sigma}_{\Tilde{w}}$ to collectively denote the tuples with components $\bar{\sigma}_{w_i}$,  $\bar{\sigma}_{x_{0,i}}$ and $\bar{\sigma}_{\Tilde{w}_i}$, respectively.
We parameterize the covariance matrix $\Sigma_{\theta}$ using a lower-triangular matrix $L_{\theta}$ with positive-valued diagonal entries, i.e., the Cholesky decomposition of the covariance matrix $\Sigma_{\theta} = L_{\theta}L_{\theta}^T$. 
We optimize over the unconstrained diagonal $\bar{l}_{i,i}$ defined through:
$$
l_{i,i} = \operatorname{softplus}\left(\bar{l}_{i,i}\right).
$$
Let $\bar{l}$ be the entire unconstrained diagonal.
Denote by $L_\theta^-$ the lower triangular part of $L_\theta$.
The variational parameters in the above guide choice are:
\begin{align*}
    \phi =\left(
    m_w, \bar{\sigma}_w
    \right), 
    \psi =\left(
    m_{x_0}, \bar{\sigma}_{x_0}, m_{\theta}, L_\theta^-, \bar{l}
    \right), 
    \text{and} \
    \Tilde{\phi} =\left(
    m_{\Tilde{w}}, \bar{\sigma}_{\Tilde{w}}
    \right).
\end{align*}

Since our guides follow normal distributions, we can easily devise fixed base distributions and deterministic transform maps that satisfy the right property.
Assume that the $\operatorname{softplus}$ function can be applied elementwise to a vector and let $\odot$ denote the elementwise product between two vectors.
Let $\epsilon$, $\eta_{x_0}$, $\eta_{\theta}$ and $\Tilde{\epsilon}$ be standard normal with the right dimension, then the random variables
$$
w = g_{\phi}\left(\epsilon\right) := m_w + \operatorname{softplus}\left(\bar{\sigma}_{w}\right)\odot \epsilon,
$$
$$
x_0= g_{\psi}\left(\eta_{x_0}\right) := m_{x_0} + \operatorname{softplus}\left(\bar{\sigma}_{x_0}\right)\odot \eta_{x_0},
$$
$$
\theta = g_{\psi}\left(\eta_{\theta}\right) := m_{\theta} + L_{\theta}\eta_{\theta},
$$
and
$$
\Tilde{w} = g_{\Tilde{\phi}}\left(\Tilde{\epsilon}\right) := m_{\Tilde{w}} + \operatorname{softplus}\left(\bar{\sigma}_{\Tilde{w}}\right)\odot \Tilde{\epsilon}
$$
follow $q_\phi(w)$, $q_\psi(x_0)$,  $q_\psi(\theta)$ and $q_{\Tilde{\phi}}(\wpp)$ respectively.
To recover the notation in Algorithms~\ref{alg:vi_prior} and~\ref{alg:vi_post},
 we  collectively write $\eta =\left(\eta_{x_0}, \eta_{\theta}\right)$ and $\lambda= g_{\psi}\left(\eta\right)$.

\subsection{Inferring the hyper parameters}
The two hyperparameters hidden in the ELBO~\qref{elbo_post_no_t} are $\beta$ and the measurement standard deviations $\sigma_y$.
Recall that $\beta$ quantifies our trust in the physical model.
We will find a point estimate of $\beta$ and use variational inference to approximate the posterior probability density of the measurement noise standard deviations $\sigma_y$.

The only fundamental difference between $\theta$ and $\sigma_y$ is that the partition function does not depend on the latter.
This simplifies a little bit the required gradients, but the overall strategy is mathematically equivalent if we include $\sigma_y$ in $\lambda$, i.e., if we assume that $\lambda = (x_0, \theta, \sigma_y)$.
Of course, we need a prior and a suitable guide.
We use the half-normal distribution as the prior $p(\sigma_{y,i})$ for each component $\sigma_{y, i}$.
The guide $q(\sigma_y)$ factorizes over each component, and $q(\sigma_{y,i})$ is chosen to be a log-normal.

Since $\beta$ is positive, we parameterize it as the softplus transformation of an unconstrained variable.
We follow the paradigm of \citep{kingma2013auto} to jointly optimize the guides and $\beta$ by maximizing the ELBO.
The only modification needed in Algorithm~\ref{alg:vi_post} is to include the gradients with respect to $\beta$ and the corresponding ADAM update.

%% file: section-numerical_examples.tex
\section{Numerical examples}

We demonstrate the efficacy of our methodology through a series of synthetic examples with increasing complexity.
In the first example, we compare IFT with the standard Bayesian approach to estimate the posterior of dynamic model parameters.
Next, we evaluate the performance of IFT against the sequential Monte Carlo square (SMC$^2$) method in the context of both state and parameter estimation. 
Lastly, we examine how IFT performs when faced with limited measurement data
and compare it with B-PINNs.
In all subsequent figures displaying the posterior distribution, we present the 5th to 95th percentile result.
We will publish the code for reproducing all the following numerical results at: \url{https://github.com/PredictiveScienceLab/pift-vi-paper-2023}.

\subsection{Parameter estimation: IFT vs. Bayesian MCMC}
\label{section_para_estiamtion}

We compare the IFT method with the standard Bayesian approach applied to a parameter estimation problem. The standard Bayesian approach aims to solve the parameter posterior $p(\theta, x_0|y)\propto p(y|\theta, x_0)p(\theta, x_0)$ using MCMC sampling. It is important to note that in order to evaluate the likelihood $p(y|\theta, x_0)$ for a given set of parameters and initial states $(\theta, x_0)$, we need to solve our ODEs numerically and compute the likelihood based on the ODEs solution.
In the following analysis, we will first demonstrate numerically that the IFT estimation result approaches the result obtained from the standard Bayesian approach as the parameter $\beta$ increases, assuming our ODE model is correct. This highlights the convergence of IFT towards the standard Bayesian approach.
Furthermore, we show that as we extend the time interval $\left[0, T\right]$, the posterior results of IFT exhibit good agreement with those obtained from Bayesian MCMC. Notably, IFT achieves this alignment while requiring significantly less time than Bayesian MCMC, making it more efficient for larger time intervals. This comparison underscores the effectiveness of IFT in handling extended time intervals.
Lastly, we showcase the ability of the IFT method to quantify model form errors using the hyperparameter $\beta$. In contrast, Bayesian MCMC converges to the maximum likelihood estimator without detecting or accounting for underlying model form errors. IFT's capability in quantifying such errors adds an additional advantage over the Bayesian MCMC approach.

We present a comparison using a three-species Lotka-Volterra model \cite{meiss2007differential} with population measurements for each species, perturbed by scaled white noise processes $\sigma_y V(t)$:
\begin{equation}
    \begin{aligned}
        \dot{x}_1(t) &= ax_1(t) - bx_1(t)x_2(t),
    \\
    \dot{x}_2(t) &= bx_1(t)x_2(t) - cx_2(t)- dx_2(t)x_3(t),
    \\
    \dot{x}_3(t)& = dx_2(t)x_3(t)- ex_3(t),
    \\
    Y(t) &= x(t) + \sigma_{y}V(t).
    \end{aligned}
    \label{lotka}
\end{equation}

The Lotka-Volterra model describes the population dynamics of three interacting species: a resource $x_1(t)$, a consumer $x_2(t)$, and a predator $x_3(t)$. The model considers the influence of species competition.
In this model, all five parameters $(a, b, c, d, e)$ are positive. As a result, we assume that the resource population $x_1(t)$ exhibits positive self-species growth, while the consumer $x_2(t)$ and predator $x_3(t)$ populations experience negative self-species growth. The interactions between adjacent species are represented by bilinear terms: $x_1(t)x_2(t)$ and $x_2(t)x_3(t)$.
We can observe the population of each species perturbed by white noise processes.

The reference (true) parameters are $(a, b, c, d, e)=(0.1, 0.02, 0.1, 0.02, 0.1)$. To improve numerical stability, it is prudent to nondimensionalize the system. 
Specifically, divide the state variables, parameters, and measurement data by $(\bar{x}_1, \bar{x}_2, \bar{x}_3) = (30, 15, 12)$, $(\bar{a}, \bar{b}, \bar{c}, \bar{d}, .\bar{e})=(0.2, 0.02, 0.2, 0.03, 0.1)$, and $(\bar{y}_1, \bar{y}_2, \bar{y}_3)=(30, 15, 12)$.
In addition, one can also normalize the time variable by the terminal time $T$.

To collect the synthetic data, we performed a $T$-second forward simulation using the 5th order Runge-Kutta method with a 0.5 second time step.
We set the initial condition $(x_1(0), x_2(0), x_3(0))^T = (10, 10, 10)^T$
and collected measurement data every 0.5 seconds.
We added white noise with a standard deviation equal to 5\% of the measurement normalizing constants $\bar{y}$.

For all numerical examples using the Lotka-Volterra model, we parameterize the time evolution state function $\hat{x}(t; w, x_0)$ using the truncated Fourier series with the 
$1+2N$
basis vectors:
\begin{align}
\psi(t)
 =
 \begin{bmatrix}
 1&
 \cos{\left(\frac{2\pi  t}{T_{\text{max}}}\right)}&
 \sin{\left(\frac{2\pi  t}{T_{\text{max}}}\right)}&
 \cdots&
 \cos{\left(\frac{2\pi N t}{T_{\text{max}}}\right)}&
 \sin{\left(\frac{2\pi N t}{T_{\text{max}}}\right)}
 \end{bmatrix}^{T},
 \nonumber
 \end{align}
 We select $T_{\text{max}}$ to be greater than $T$ because the parameterization is periodic whereas the dynamical system response does not have to be.
 We choose the constant basis $\psi_0 (t)=1$ as the dependent basis and solve \qref{wj}.
 This can be analytically simplified for the truncated Fourier series (details in \ref{appendix:fourier}).

We use
$N=20$ Fourier terms to parameterize the prior fields.
Since the population and the five parameters are positive, we parameterize them as the exponential or softplus transformation of the unconstrained variables. 
The prior distribution of each nondimensionalized unconstrained initial state follows a standard normal distribution with a mean of 0 and a standard deviation of 1. 
The five nondimensionalized unconstrained parameters also have independent standard normal priors.

Regarding the guide, we use two separate diagonal multivariate Gaussian distributions to approximate the posterior of the Fourier coefficients and the unconstrained initial states.  
A full-rank multivariate Gaussian distribution approximates the posterior of the five unconstrained dynamical system parameters.
We initialized the variational parameters
$\phi=\left(m_w, \bar{\sigma}_w\right)=(0, -4)$, 
$\psi=\left(m_{x_0}, \bar{\sigma}_{x_0}, m_{\theta}, L^-_{\theta}, \bar{l}\right)=(0, -4, 0, 0, -1)$, and $\Tilde{\phi}=\left(m_{\wpp}, \bar{\sigma}_{\wpp}\right)=(0, -4)$.
Notice that the initial values of the standard deviations for the Fourier coefficients and initial states are small.
Our experiments showed that this choice can accelerate the convergence of the algorithm.

To optimize the ELBO, 
we subsample the data with a batch size of $m_d = 10$.
The other sample sizes in Algorithm \ref{alg:vi_post} are $\left(n_{\epsilon},n_t, n_{\tilde{w}}, \tilde{n}_\epsilon, \tilde{n}_t\right)=(1, 10, 1, 1, 10)$.
We performed 200,000 steps of optimization using Adam optimizer~\cite{kingma2014adam} with an initial learning rate of 0.001.
An exponential learning rate scheduler decays the learning rate to 0.0001 at the final iteration.
We also experimented with KL annealing (first 20,000 iterations gradually bring in the term of the ELBO that requires sampling from the prior), albeit we did not observe a significant improvement.

In terms of the standard Bayesian MCMC approach, we use No-U-Turn Sampler (NUTs) \cite{hoffman2014no} in Numpyro (\cite{phan2019composable}, \cite{bingham2019pyro}) to sample from the posterior.
We did 1,000 warm-up steps followed by 5,000 samples.

\subsubsection{Correct model with increasing $\beta$ and fixed $T$}
\begin{figure}[!h]
    \centering
    \includegraphics[scale=.42]{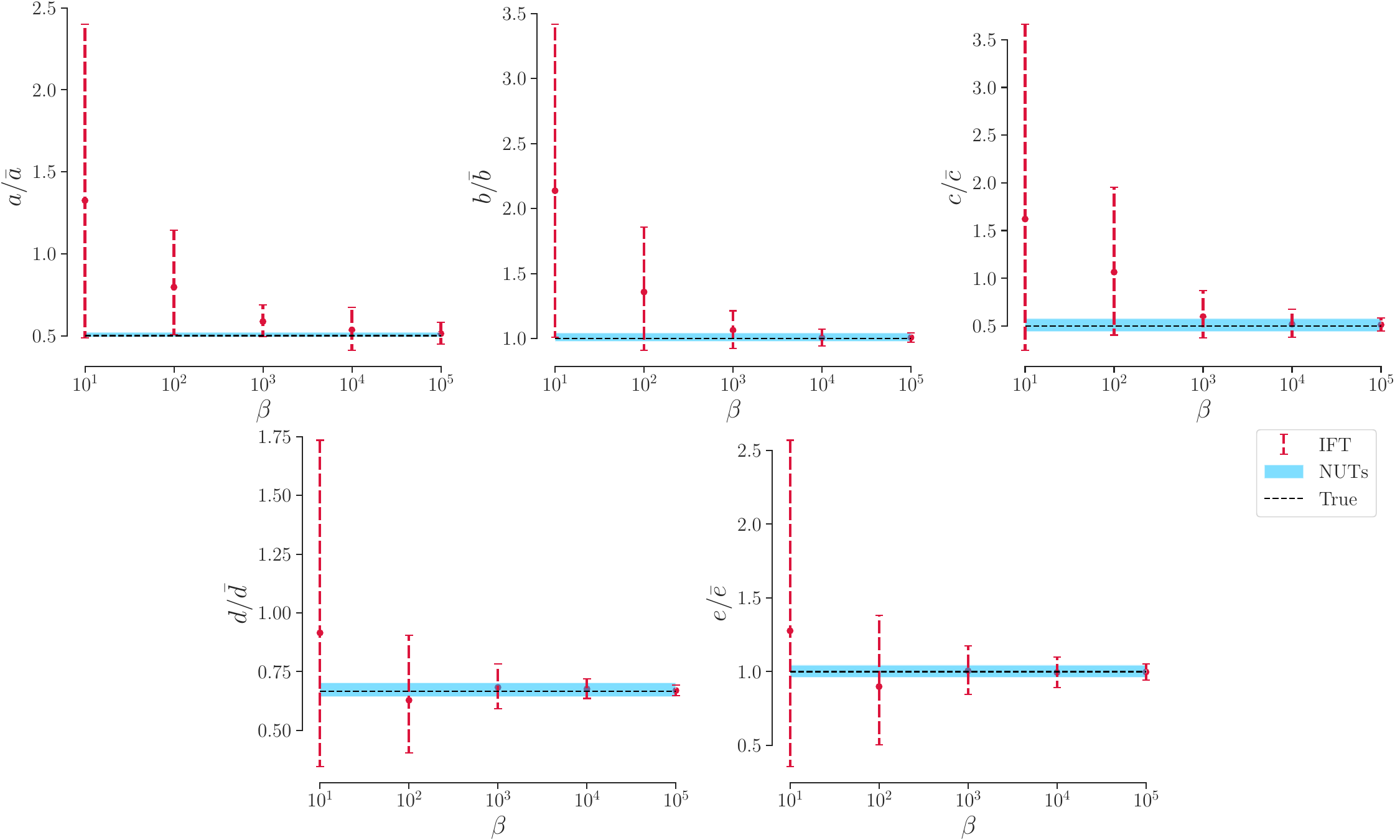}
    \caption{IFT parameter posteriors with increasing $\beta$ and NUTs parameter posterior.}
\label{pift_converge_NUTs}
\end{figure}

Fig. \ref{pift_converge_NUTs} depicts the posterior distribution of the five parameters.
In this example, we generated 50-second measurement data. 
The blue lines show the NUTs posterior results.
For the IFT method, we solved the IFT problem five times, gradually increasing the value of $\beta$ from 10 to 100,000. 
Remarkably, we observe that as we progressively raise $\beta$, the IFT estimation consistently converges towards the NUTs result. 

\subsubsection{Correct model with fixed $\beta$ and increasing $T$ }
\begin{figure}[!h]
    \centering
    \includegraphics[scale=.42]{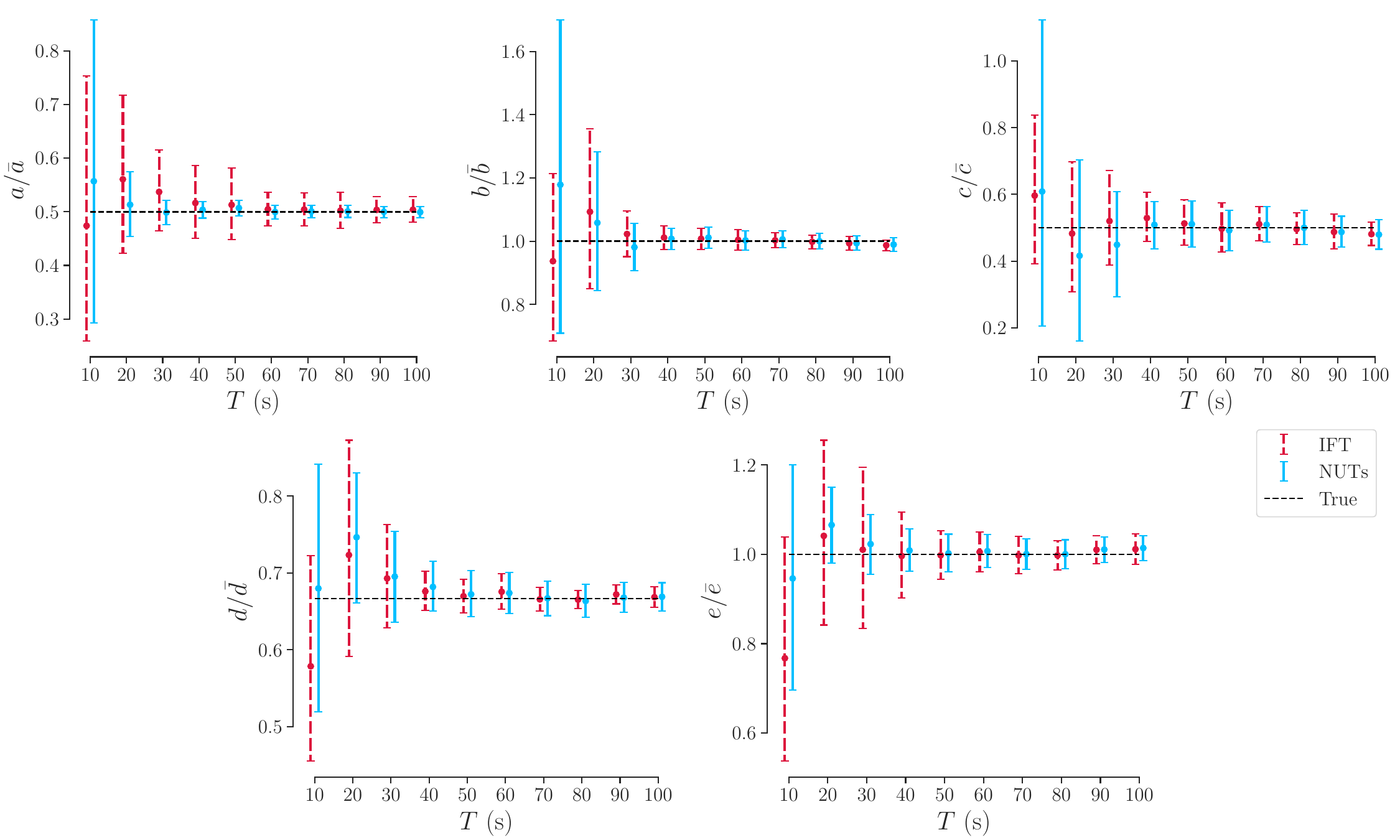}
    \caption{IFT and NUTs parameter posteriors with increasing time interval.}
\label{pift_NUTs_vary_time}
\end{figure}

In Fig. \ref{pift_NUTs_vary_time}, we compare the posterior results obtained using the IFT and NUTs methods for the five parameters as the time interval $[0, T]$ increases. For this analysis, we set $\beta$ to a fixed value of 100,000 in the IFT method.
The figure shows a strong alignment between the IFT and NUTs results as $T$ increases. 
This demonstrates that the IFT method can capture the posterior distribution across longer time intervals, similar to the NUTs method.
However, it is important to note that IFT offers a significant advantage over NUTs when considering larger values of $T$. As shown in Fig. \ref{comptime}, which summarizes the computational times required for the results presented in Fig. \ref{pift_NUTs_vary_time}, the computational time for IFT remains constant since we use the same number of Fourier bases for all $T$. On the other hand, NUTs exhibits a substantial increase in computational time for larger $T$.
Therefore, the IFT method provides a more scalable and efficient approach compared to NUTs when dealing with extended time intervals.

\subsubsection{Model form error}
\begin{figure}[!htpb]
    \centering
    \includegraphics[scale=.42]{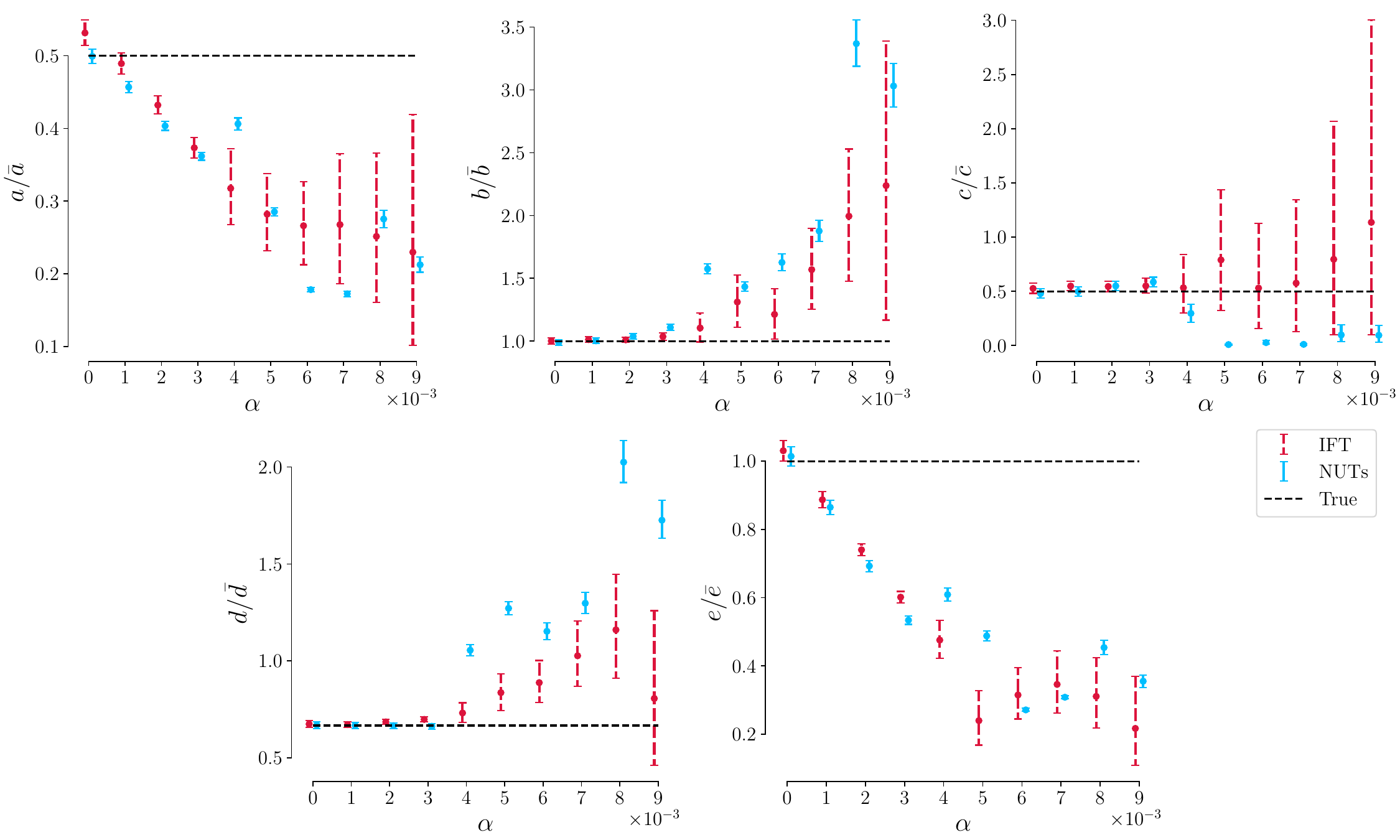}
    \caption{IFT and NUTs parameter posteriors with increasing model form error.}
\label{pift_model_error}
\end{figure}

\begin{figure}[!htpb]
    \centering
    \includegraphics[scale=.5]{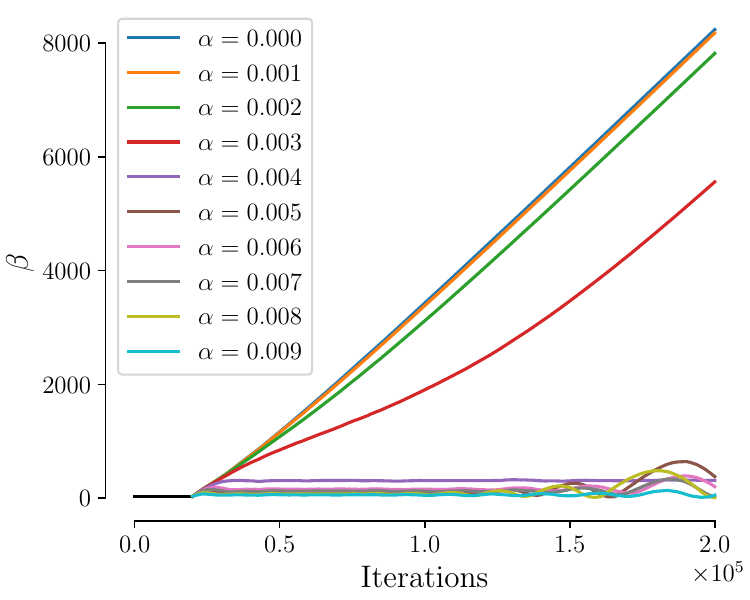}
    \caption{IFT hyperparameter $\beta$ training iterations for different model form error levels.}
\label{pift_beta_model_error}
\end{figure}

Next, we demonstrate how IFT works when our model has unknown errors. 
Specifically, we assume the correct model is 
\begin{align}
    \dot{x}_1(t) &= ax_1(t) - bx_1(t)x_2(t) - \alpha x_1(t)x_3(t),
    \nonumber
    \\
    \dot{x}_2(t) &= bx_1(t)x_2(t) - cx_2(t)- dx_2(t)x_3(t),
    \nonumber
    \\
    \dot{x}_3(t)& = dx_2(t)x_3(t)- ex_3(t) + \alpha x_1(t)x_3(t),
    \nonumber
\end{align}
where we include the bilinear term $\alpha x_1(t)x_3(t)$ between the resource and predator.
Increasing the value of $\alpha$
indicates a larger model form error.
When $\alpha=0$, our model is perfect.
We used this model to generate the ground truth measurement data, but still applied IFT using the original model in \qref{lotka}.
To allow IFT to quantify the model form errors, we learned $\beta$ starting from an initial value of 1.
This time, we generated 100-second data to calibrate the parameters.
We ran NUTs for each $\alpha$ to sample from the parameter posterior using the incorrect model.
Fig. \ref{pift_model_error} presents the posterior distributions of the five parameters, while Fig. \ref{pift_beta_model_error} depicts the training iterations of the IFT hyperparameter $\beta$ for different values of $\alpha$.
As the model form error, represented by $\alpha$, increases, we observe that the IFT hyperparameter $\beta$ tends to approach smaller values, and the parameter posteriors exhibit greater uncertainties. In contrast, NUTs, which is not aware of the underlying model form error, still produces confident posteriors.

\subsection{State and parameter estimation: IFT vs. particle filter}

In this section, we highlight the effectiveness of the IFT method in solving the challenging batch estimation problem, which involves reconstructing the joint posterior distribution of the entire state trajectory $x$ from time 0 to $T$ and the parameter $\theta$, given all available data $y$ up to time $T$.
Traditionally, to simplify the problem, researchers discretize the dynamic model and focus on estimating the joint posterior of the discretized states and parameters. However, even with this simplification, the batch estimation problem quickly becomes computationally heavy and can be intractable.
As alternative approaches, filtering methods are often applied to estimate the marginal posterior distribution of states given the measurement data up to the current time $p(x_{t}|y_{1:t})$, and smoothing methods are used to estimate the marginal posterior distribution of states given all the measurement data $p(x_t|y_{1: T})$.
In this comparison, we specifically compare IFT with the particle filter technique, which is a powerful method commonly used for state estimation. However, it's important to note that standard particle filters focus only on state estimation and do not address parameter estimation. To provide a fair comparison, we employ the SMC$^2$ method developed by \citet{chopin2013smc2}, which allows for joint estimation of both states and parameters.
We should not expect the posteriors from IFT and SMC$^2$ to match with each other exactly since they solve batch and filtering problems, respectively.

We investigate the same Lotka-Volterra model in \qref{lotka}. To apply SMC$^2$, we need to transform the deterministic model into an SDE and discretize it using the Euler-Maruyama scheme.
According to \ref{appendix:sde}, we set the diffusion constant in the SDE to $(2\beta)^{-1/2}$ for SMC$^2$. This ensures consistency with the IFT method and facilitates a meaningful comparison between the two approaches.
For both IFT and SMC$^2$, we employ the same prior distributions for the initial states and parameters. However, particle filters like SMC$^2$ require a properly initialized set of particles drawn from the prior distribution. The initial state prior used in section~\ref{section_para_estiamtion} is too spread-out, which leads to a low effective number of particles for SMC$^2$. To ensure fairness in the comparison, we use a more informative Lognormal([-1, -0.5, -0.25], [-0.25, -0.2, -0.2]) prior for the three states, which is more suitable for SMC$^2$.
The optimization settings for IFT are identical to those described in section \ref{section_para_estiamtion}.
In the case of SMC$^2$, we utilize the Pyfilter Python package.  We employ 500 particles to estimate the parameters, and for each parameter particle, there are 1,000 particles used to estimate the states. The state particles are updated using the auxiliary particle technique filter~\cite{pitt1999filtering}.

\subsubsection{Correct model with increasing $\beta$ and fixed $T$}
\begin{figure}[!htpb]
    \includegraphics[scale=.42]{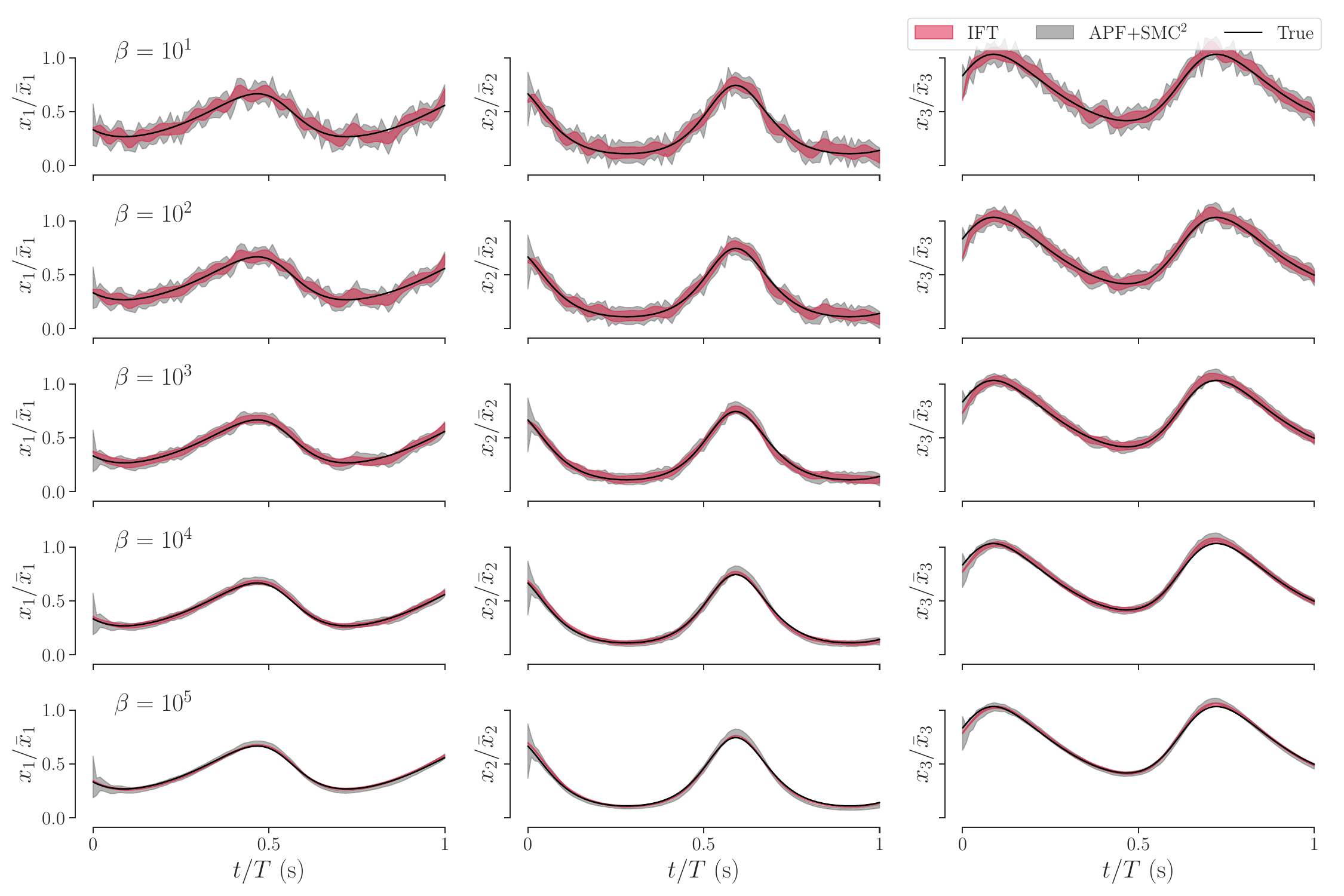}
    \caption{IFT and SMC$^2$ state posteriors with increasing $\beta$.}
\label{pift_smc_state}
\vspace{1cm}
    \includegraphics[scale=.42]{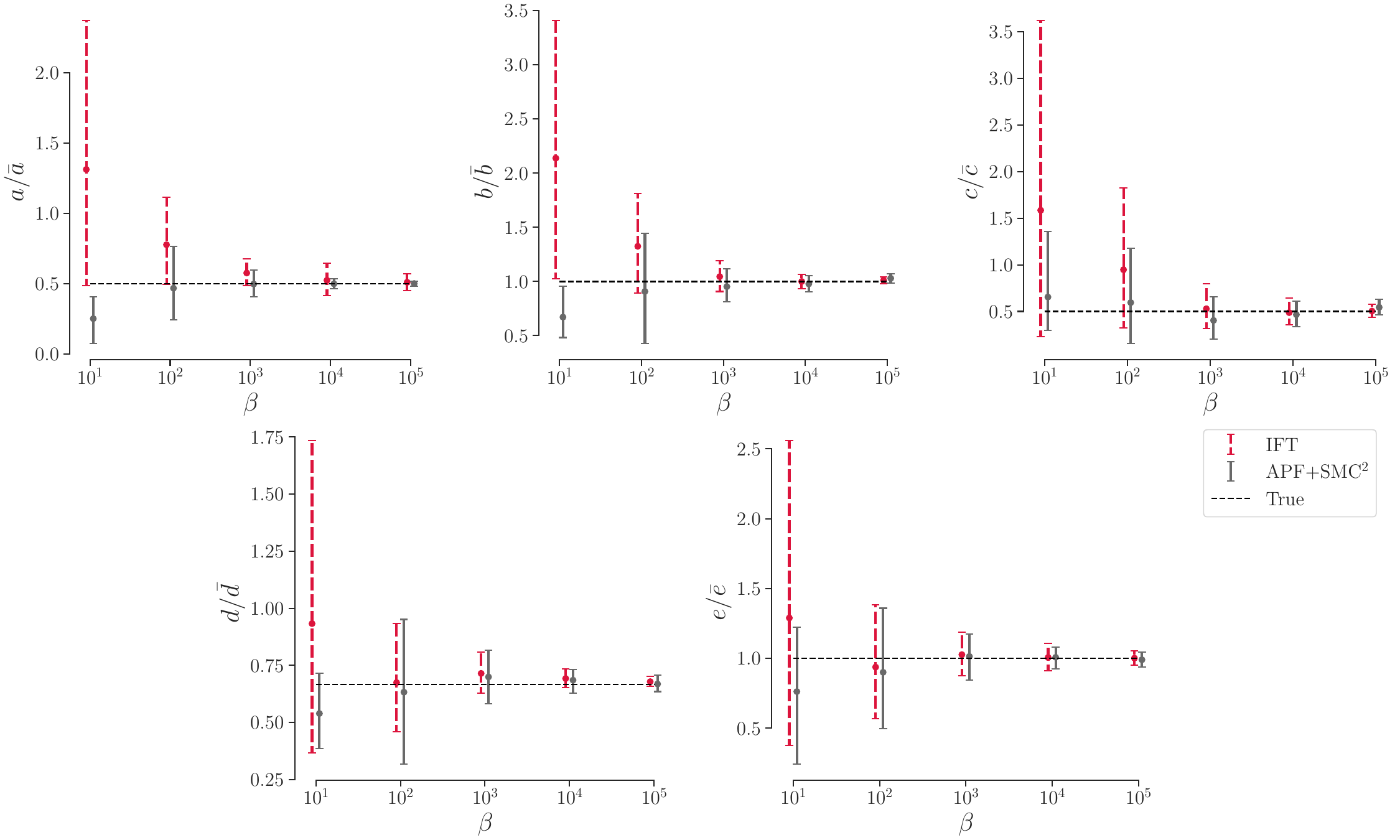}
    \caption{IFT and SMC$^2$ parameter posteriors with increasing $\beta$.}
\label{pift_smc_para}
\end{figure}

In Fig. \ref{pift_smc_state}, we present the state posterior distributions obtained from IFT and SMC$^2$ when the model is correctly specified. 
The measurement data spans 50-seconds.
To examine the effect of $\beta$, we vary its value along the vertical axis.
Overall, we observe that the state posterior uncertainty obtained from IFT is generally lower than that of SMC$^2$. This is expected as IFT utilizes all available measurement data. 
In contrast, SMC$^2$ performs filtering estimation and only considers the measurement data up to the current time.
Fig. \ref{pift_smc_para} illustrates the parameter posterior distributions obtained from both IFT and SMC$^2$. As $\beta$ increases, we observe a better agreement between the two methods.

\subsubsection{Correct model with fixed $\beta$ and increasing $T$}

\begin{figure}[!htpb]
    \includegraphics[scale=.42]{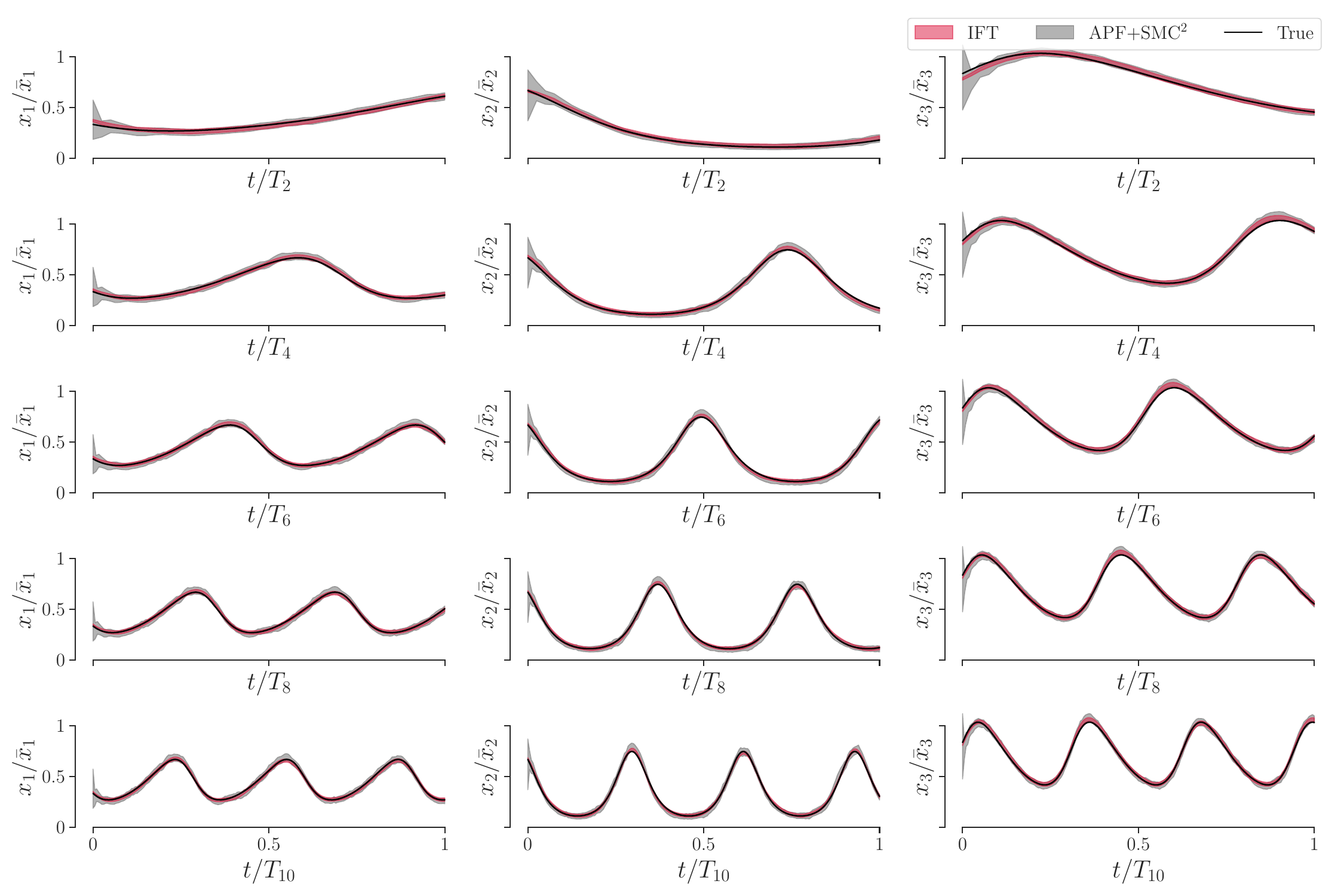}
    \caption{IFT and SMC$^2$ state posteriors with increasing $T$.}
\label{pift_smc_stat_vary_time}
\vspace{1cm}
    \includegraphics[scale=.42]{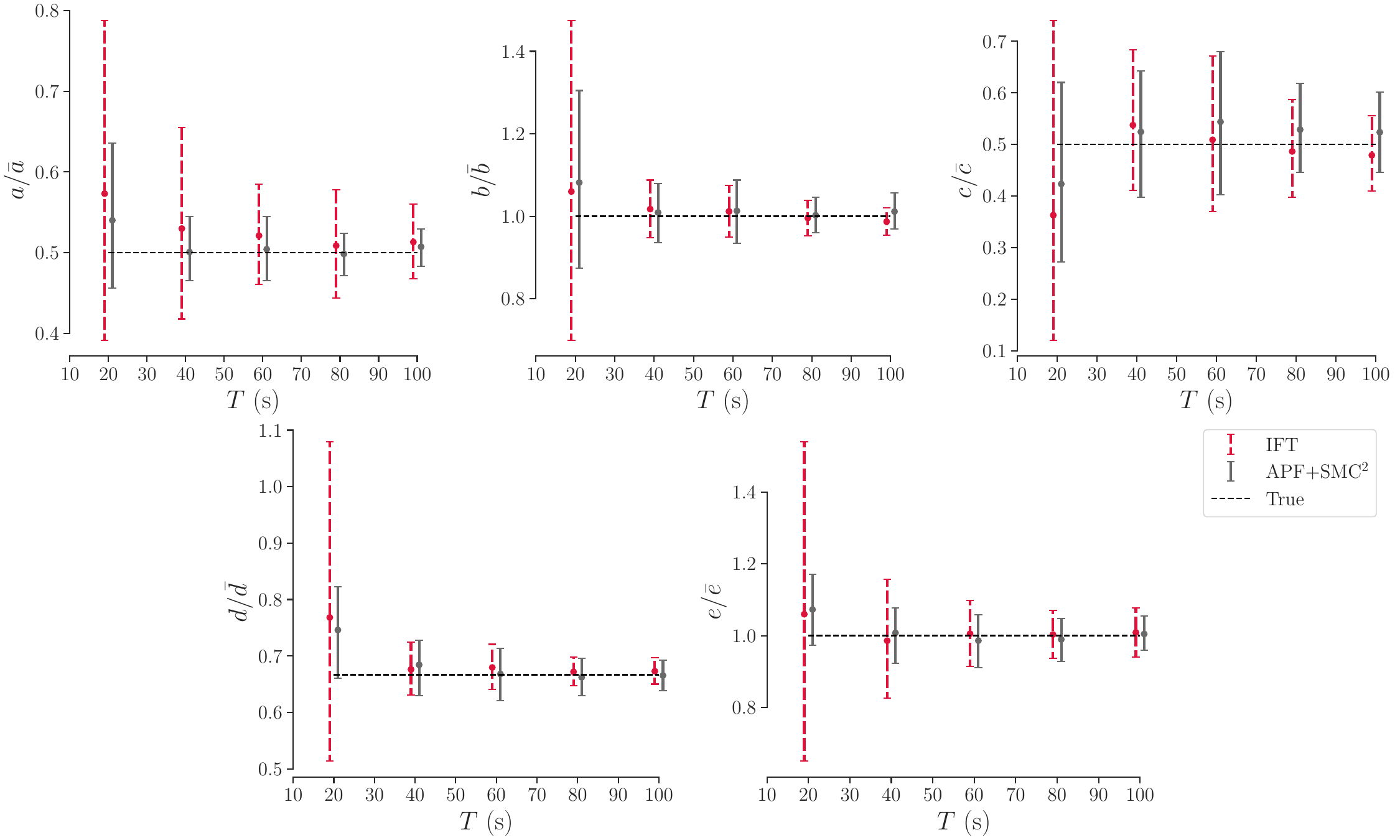}
    \caption{IFT and SMC$^2$ parameter posteriors with increasing $T$.}
\label{pift_smc_para_vary_time}
\end{figure}

\begin{figure}[!htpb]
    \centering
    \includegraphics[scale=.5]{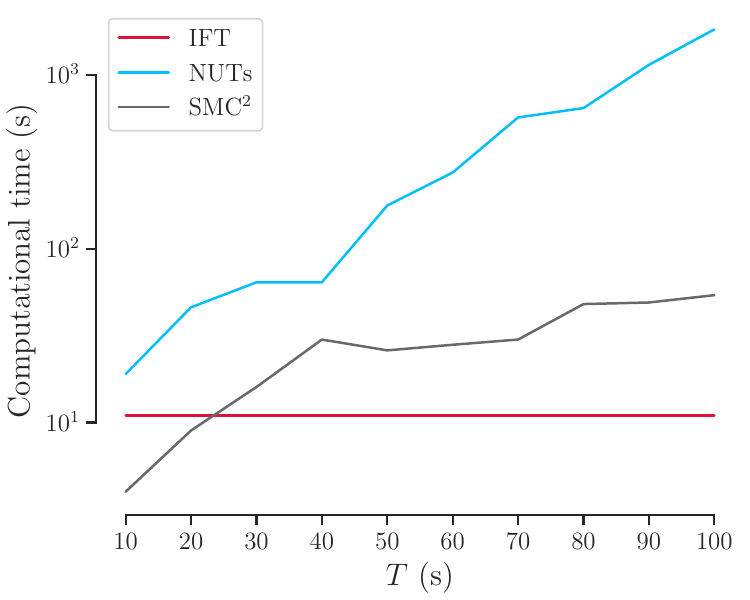}
    \caption{Comparison of computational time.}
\label{comptime}
\end{figure}

We examine the state and parameter estimation results with a fixed $\beta=10,000$ while increasing the time intervals $[0, T]$ from 10s to 100s.
The corresponding results are presented in Figs. \ref{pift_smc_stat_vary_time} and \ref{pift_smc_para_vary_time}.
In general, both IFT and SMC$^2$ demonstrate similar performance as the time interval $T$ increases. 
However, when considering computational efficiency, IFT outperforms SMC$^2$ as shown in Fig. \ref{comptime}. With longer time intervals, IFT exhibits superior scalability in terms of computational time. This advantage makes IFT particularly well-suited for handling larger time intervals, providing significant practical benefits.

The final remark is that in the case of model form errors, it is possible to learn the hyperparameter $\beta$ in SMC$^2$ by treating it as a model parameter. However, it should be noted that in this context, $\beta$ does not have the interpretation of quantifying model form errors but rather represents the volatility of the system.
This differs from the role it plays in IFT, where it explicitly quantifies model form errors.

\subsection{Posterior and posterior predictive estimation with limited measurement data}

We demonstrate how IFT works with limited measurement data.
Let us examine the Duffing oscillator as our final numerical example.
It is described by the following set of equations:
\begin{align}
     \dot{x}_{1}(t) &= x_2 (t), \nonumber
     \\
     \dot{x}_{2}(t) &= -\delta x_{2}(t) - \alpha x_{1}(t) - \rho x_{1}(t)^3 + \gamma \cos{(\omega t)}, \nonumber
     \\
     Y(t) &= x_1 (t) + \sigma_y V(t).\nonumber
 \end{align}
 Such a system represents a damped oscillator with a nonlinear restoring force term, $\rho x_1 (t)^3$. In this case, we will utilize only the position measurement to estimate states and parameters.

Our inference problem focuses on estimating the time-evolution state trajectories, namely $x_1(t)$ and $x_2(t)$, as well as the parameters $\delta$, $\alpha$, and $\rho$.
For the driving force, we know that the amplitude is $\gamma = 0.37$ m, and the frequency is $\omega = 1.2$ rad/s.
In this scenario, we choose the reference parameter values of $\delta = 0.3$, $\alpha = -1$, and $\rho = 1$.
Similar to the first example, we should work with the nondimensionalized system.
The normalization constants are $(\bar{x}_1, \bar{x}_2) = (1.5, 1)$, $(\bar{\delta}, \bar{\alpha}, \bar{\rho}) = (1, 1, 1)$, and $(\bar{y}) = (1.5)$.

We generated synthetic measurement data for a duration of 50 seconds, using a discretized time step of 0.01s. 
The measurement noise standard deviations are set to 5\% of the measurement normalization constant. 
To reduce the data size, we downsampled the measurements to 10\%, which means we retained data points at intervals of 0.1s.
The initial states are $(x_1(0), x_2(0))^T = (1, 0)^T$.
In this example, we used the radial basis function 
$
\psi_{k}(t) 
=
\exp\left\{-\frac{(x-z_k)^2}{2\sigma_{\psi, k}^2}\right\}
$
to parameterize the state function.
Specifically, there are $d_w=d_x\times K = 2\times 100$ radial basis terms,
whose centers $z_{k}$ are evenly spaced from 0 to 1 and length scales $\sigma_{\psi, k}$ are all 0.02.
We did not pose the positive constraint for states and parameters.
This time we only train 100,000 iterations and $\beta=200$.
The other optimization setups are identical to the first example.

\begin{figure}[!htpb]
    \includegraphics[scale=.32]{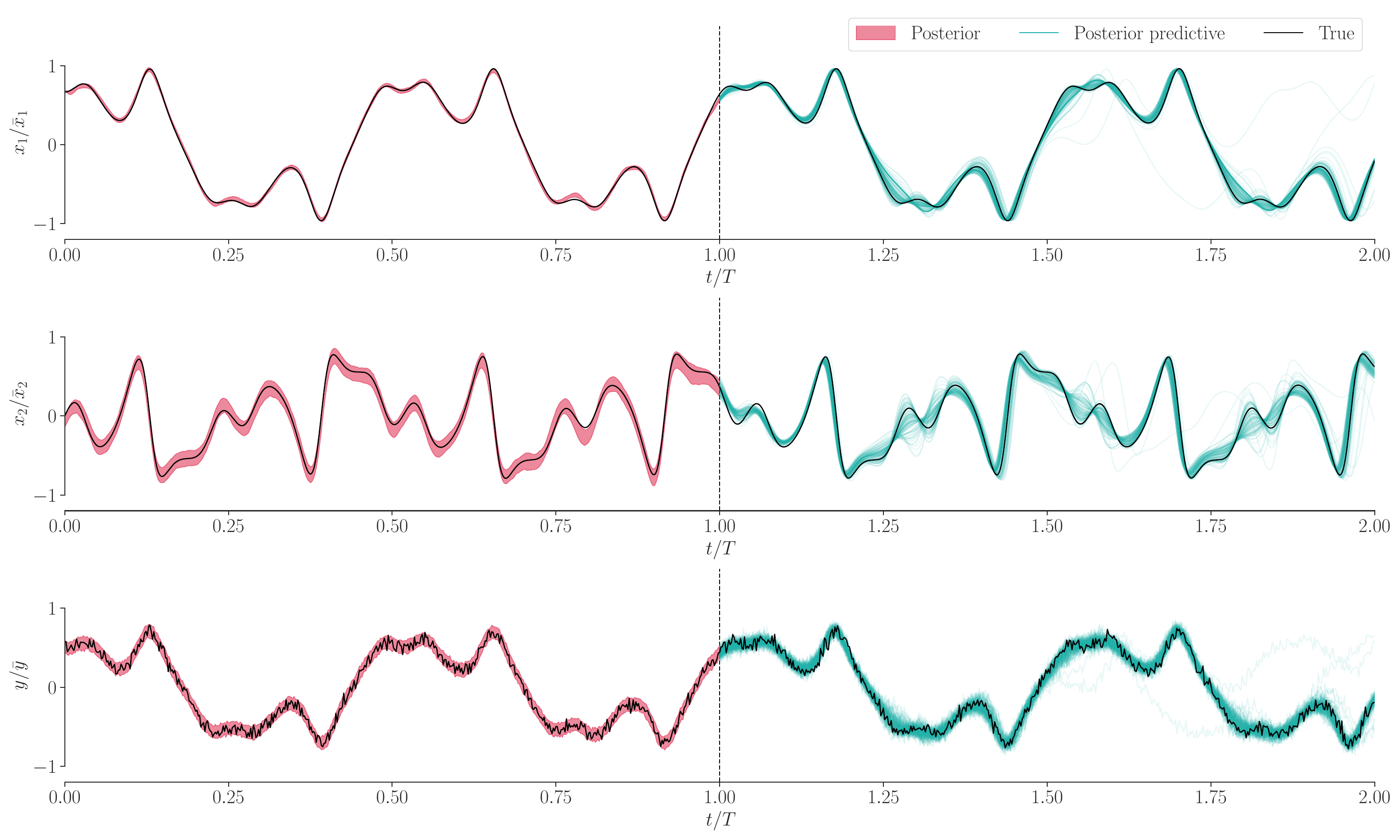}
    \caption{
    IFT: posterior and posterior predictive distributions of Duffing oscillator.
    }
\label{post_pred_do}
\vspace{1cm}
    \includegraphics[scale=.42]{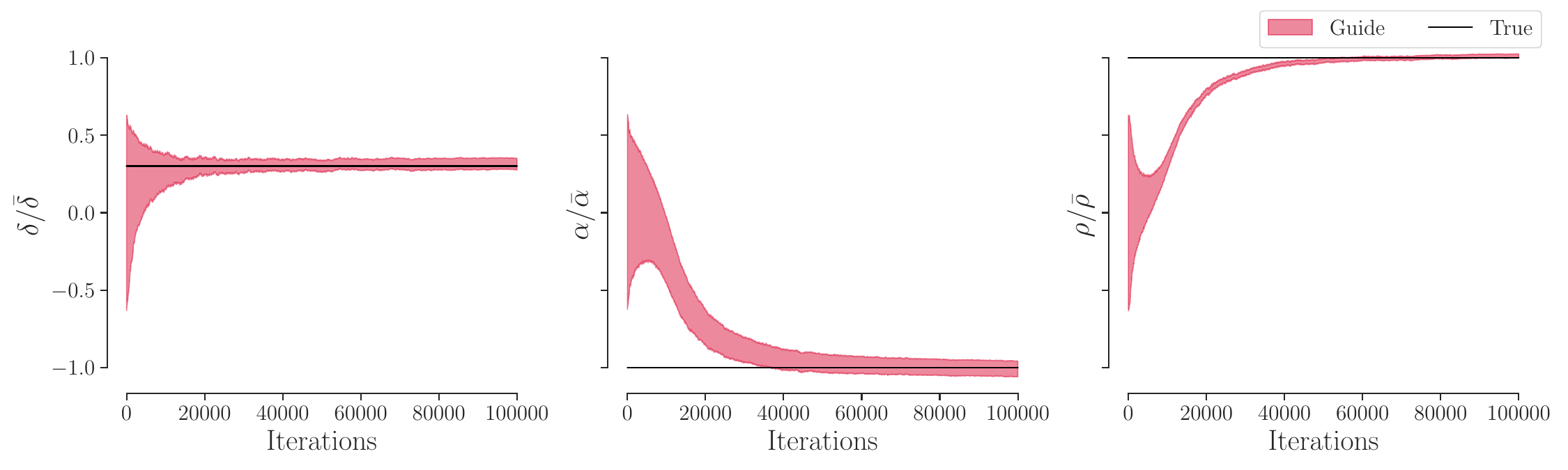}
    \caption{
    IFT: parameter guide iterations.}
\label{para_guide_iter}
\end{figure}

Fig.~\ref{post_pred_do} depicts the results of state estimation.
The first two rows correspond to the states.
The last row corresponds to the observations.
The figure is split in half.
The left half reconstructs the state using the available data (red color for the posteriors) whereas the right half makes future predictions (green color for the predictions).
Specifically, we show 100 samples for the future predictions.
Notice that since we only have access to the position measurement, the uncertainty in estimating the velocity, $x_2$, is relatively high.
Overall, the reconstruction matches very well the true trajectories shown in black.

Fig.~\ref{para_guide_iter} illustrates the result of parameter identification.
We see how the guide converges to a region about the true parameters as the number of iterations increases.

\subsection{
Comparison to B-PINNs
}
\label{sec:compare-to-b-pinns}

We compare IFT to B-PINNs using the same Duffing oscillator example.
We numerically demonstrate that neglecting the IFT normalization constant identifies suboptimal parameters.

To give a fair comparison, all the experimental setups of B-PINNs are the same as IFT's setup, except that we specify a standard normal prior for $w$ in B-PINNs and the fictitious residual measurement standard deviation $\sigma_{y^r}$.
For $\sigma_{y^r}$, from \qref{hwt}, we write down the estimator of the log-prior density of ITF using $n_t$ number of time samples:
\begin{align*}
     \widehat{
    \log
    p_{\beta}(w|x_0, \theta)
    }
    =
    -\frac{\beta T}{n_t}
    \sum_{i=1}^{n_t}
\left\Vert
\dot{\hat{x}}(t_i; w, x_0)-
    f(\hat{x}(t_i;w, x_0), t_i; \theta)
    \right\Vert^2
    -
    Z_\beta(x_0, \theta).
\end{align*}
Next, from \qref{logpyr}, we write down the log-likelihood of the fictitious residue measurement term in B-PINNs:
\begin{align*}
    \log 
    p(y^r=0|w, x_0, \theta)
    =
    -\frac{1}{2\sigma^2_{y^r}}
    \sum_{i=1}^{n_r}
\left\Vert 
\dot{\hat{x}}(t_i; w, x_0)-
    f(\hat{x}(t_i;w, x_0), t_i; \theta)
    \right\Vert^2
    -
    Z(N_r, \sigma_{y^r}).
\end{align*}
Comparing the two log-densities, in the numerical experiment, we randomly generate $n_r=n_t$ number of fictitious residue measurements at each B-PINNs' iteration and choose $\sigma_{y_r}$ such that 
$
\frac{1}{2\sigma^2_{y^r}}
=
\frac{\beta T}{n_t}
$.

\begin{figure}[!htpb]
    \includegraphics[scale=.32]{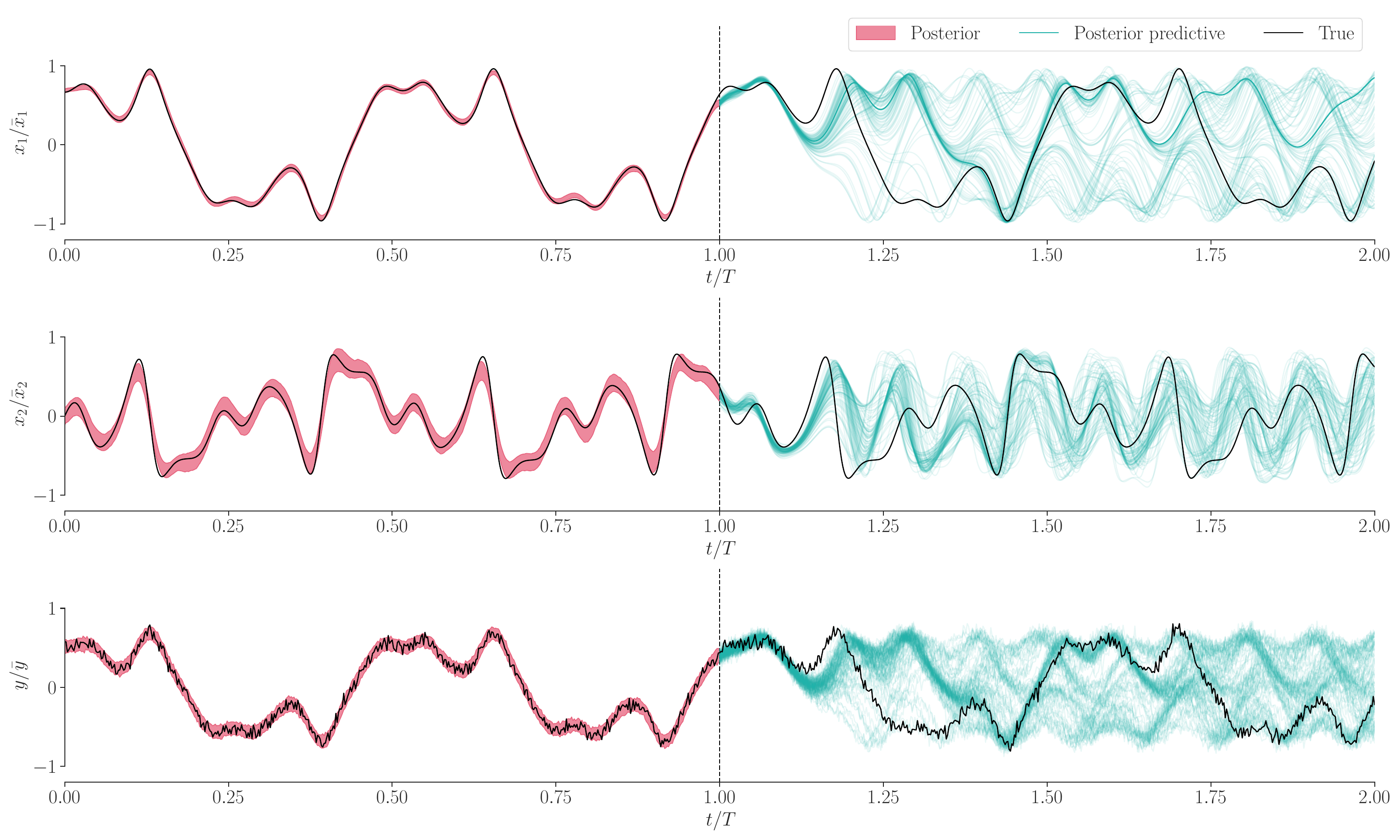}
    \caption{
    B-PINNs: Posterior and posterior predictive distributions of Duffing oscillator.
    }
\label{bpinns_postpred}
\vspace{1cm}
    \includegraphics[scale=.42]{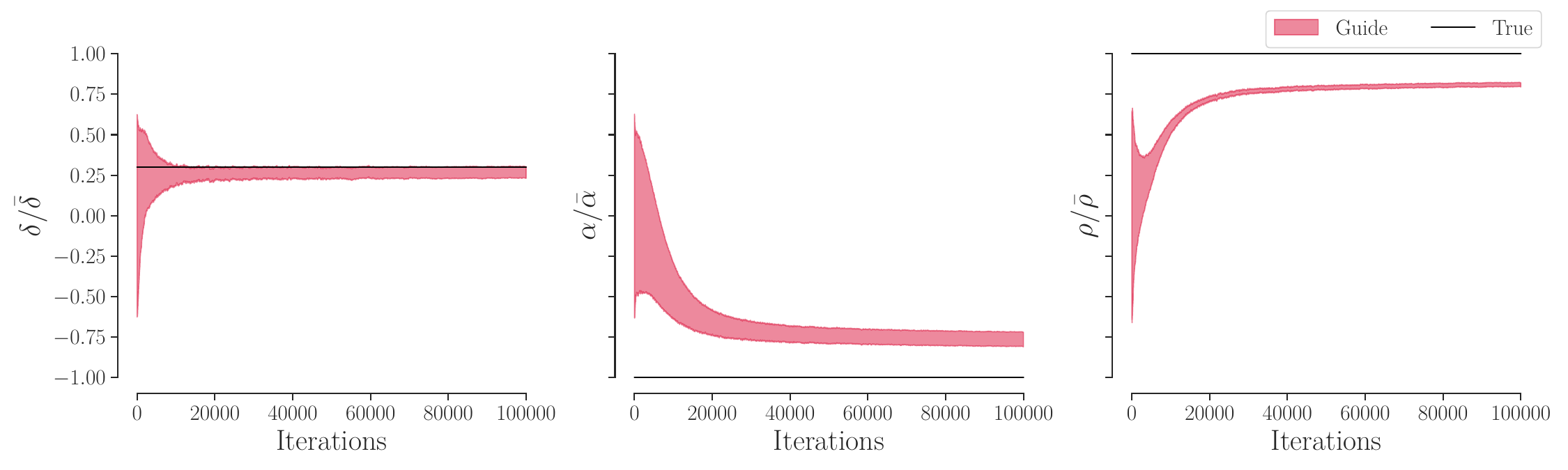}
    \caption{
    B-PINNs: Parameter guide iterations.}
\label{bpinns_para_post}
\end{figure}

Figs. \ref{bpinns_postpred} and \ref{bpinns_para_post} plot the results of B-PINNs.
We observe the posterior distribution of state functions is in agreement with IFT's result.
However, the posterior distribution of the model parameter is not close to the ground truth values, and so is the posterior predictive distribution.

%% file: section-conclusions.tex
\section{Conclusions}

We have developed a novel methodology based on physics-informed information field theory (IFT) for Bayesian estimation of states and parameters in dynamical systems. Unlike traditional approaches such as Kalman and particle filters, IFT tackles the challenging batch estimation problem by estimating continuous-time evolution states and parameters with access to all measurement data.
IFT constructs a physics-informed conditional prior through a path integral formulation, enabling us to incorporate trust in the model and quantify model form errors. We have utilized SDE theory to establish the relationship between the physics-informed conditional prior and an associated Euler-Maruyama discretized SDE. 
Furthermore, we have provided theoretical proof of the convergence of mean and variance.
To address the challenge of approximating the posterior distribution, we have developed a stochastic variational inference algorithm that can effectively handle large datasets. This algorithm differs from standard variational inference algorithms as it accounts for the intractable field partition function present in the physics-informed prior.
We have conducted numerical comparisons between IFT and Bayesian MCMC methods in parameter estimation problems. The results demonstrate that, when the model is accurate, IFT achieves posterior results comparable to Bayesian MCMC. Moreover, IFT offers the additional capability of quantifying model form errors and exhibits greater robustness in the presence of such errors.
In the context of state and parameter estimation, we have also compared IFT with SMC$^2$. Remarkably, IFT produces comparable results to SMC$^2$ while demonstrating superior scalability for large datasets.
Finally, we have demonstrated IFT's performance in situations with limited measurement data and compare the result to B-PINNs, showcasing its ability to handle challenging scenarios and provide reliable estimates.

In summary, the developed IFT methodology offers a powerful framework for Bayesian estimation in dynamical systems. It leverages physics-informed priors, scalable variational inference algorithms, and the ability to quantify model form errors. The numerical experiments highlight its effectiveness and demonstrate its potential as a valuable tool for a wide range of applications.

%% file: section-appendix.tex
\appendix

\section{
From Euler-Maruyama discretized SDEs to $p_{\beta}(w\vert x_0, \theta)$ }
\label{appendix:sde}

We connect the proposed physics-informed conditional prior $p_{\beta}(w\vert x_0, \theta)$ to a discretized SDE using the Euler-Maruyama scheme.
We claim that establishing this connection demonstrates that the prior joint probability over the function values is independent of the choice of the basis.

Let $B(t)$ be a $d_x$-dimensional Brownian motion and consider the SDE
\begin{align}
dX(t) = f(X(t), t; \theta)\ dt + (2\beta)^{-\frac{1}{2}}dB(t); 
\quad X(0) = x_0, \label{sde_prior}
\end{align}
in the It\^{o} sense.
We discretize the path from $x(0)$ to $x(T)$ using $N$ equidistant time intervals of size $\Delta t = \frac{T}{N}$. 
For notational convenience, we set
    \begin{align*}
        t_i = \frac{i}{N}T,\ 
        X_i = X(t_i),
        \text{and} \
        x_i = x(t_i)
        .
    \end{align*}

\begin{figure}[!h]
    \centering
    \begin{tikzpicture}
        \draw[thick] (0, -2) -- (0, 2) ;
         \filldraw [black] (0, 2) node[anchor=south]{$\mathbb{R}^{d_x}$};
        \draw[thick] (6, -2) --(6, 2);
        \filldraw [black] (6, 2) node[anchor=south]{$\mathbb{R}^{d_x}$};
        \filldraw [black] (0, -2) node[anchor=north]{$t_0$};
        \filldraw [black] (3, -2) node[anchor=north]{$t_i$};
        \filldraw [black] (4, -2) node[anchor=north]{$t_{i+1}$};
        \filldraw [black] (6, -2) node[anchor=north]{$t_N$};
        \draw(1, -2) -- (1, 2);
        \draw(2, -2) -- (2, 2);
        \draw(3, -2) -- (3, 2);
        \draw(4, -2) -- (4, 2);
        \draw(5, -2) -- (5, 2);
        \filldraw [black] (0, 0) circle(2pt) node[anchor=east]{$x_0$};

        \draw[thick, red] plot [smooth]  coordinates { (0,0) (1,1) (2,0.5) (3,2) (4, 0.65) (5, 0.55) (6, 0.8)};
        \draw[thick, red] plot [smooth]  coordinates { (0,0) (1,-0.3) (2, -0.2) (3, -1) (4, -0.5) (5, -1.5) (6, -0.7)};
        \filldraw [black] (3, -1) circle(2pt) node[anchor=north west]{$x_i$};
        \filldraw [black] (4, -0.5) circle(2pt) node[anchor=south west]{$x_{i+1}$};
        \filldraw [black] (6, -0.7) circle(2pt) node[anchor=west]{$x_T$};
        \draw[thick, dashed, blue] plot   coordinates { (0,0) (1,1) (2,0.5) (3,2) (4, 0.65) (5, 0.55) (6, 0.8)};
        \draw[thick, dashed,blue] plot  coordinates { (0,0) (1,-0.3) (2, -0.2) (3, -1) (4, -0.5) (5, -1.5) (6, -0.7)};
       \draw[thick,red] plot  coordinates { (0.5,3) (1.5, 3)};
       \node at (4, 3) {IFT parameterized path};
       \draw[thick, dashed,blue] plot  coordinates { (0.5,2.5) (1.5, 2.5)};
       \node at (4, 2.5) {Discretized SDE path};
    \end{tikzpicture}
    \caption{Schematic of discretized SDEs paths (blue dashed lines) and IFT parameterized random paths $\hat{x}(t; w, x_0)$ (red solid lines).}
    \label{fig:time_slicing}
\end{figure}

Use Euler-Maruyama scheme to discretize \qref{sde_prior}:
    \begin{align}
        X_{i+1} \approx X_i + f(X_i, t_i; \theta)\Delta t
        +
        \left(
        \frac{\Delta t}{2\beta}
        \right)^{\frac{1}{2}}
        V_i,
        \label{EMS}
    \end{align}
    where $V_i$ follows a multivariate standard Gaussian with identity covariance matrix.
This discretization  is approximate but becomes increasingly more accurate for smaller $\Delta t$.
Technically, the solution of the discretized version, \qref{EMS}, converges strongly to the solution of the SDE, \qref{sde_prior}.

From \qref{EMS}, we write down the probability density for all discretized states
\begin{align}
    p_{SDE}(x_{1:N}\vert x_0, \theta)
    \propto
    \exp
    \left\{
    -
        \beta
        \sum_{i=0}^{N-1}
        \Big\Vert
        \frac{x_{i+1}-x_i}{\Delta t}
         - f(x_i, t_i; \theta)
        \Big\Vert^2
        \Delta t
    \right\}
    \label{sde_px1N}
\end{align}

We define the map 
$g(\cdot; t_{1:N}, x_0):\mathbb{R}^{d_w}\to \mathbb{R}^{d_x\times N}$,
$w \mapsto x_{1:N}=\hat{x}(t_{1:N};w, x_0)$.
Here, we vectorize the discretized states $x_{1:N}$ into
$
[x_{11}, \cdots, x_{1N}, x_{21}, \cdots, x_{2N}, \cdots, x_{d_x 1}, \cdots, x_{d_x N}]
$.
Then we assume $d_w=d_x \times N$ and $g$ is bijective.
We apply the change of variables formula to define a probability density $\Tilde{p}(w)$ for $w$ from the discretized SDE density in \qref{sde_px1N}:
\begin{align}
    \tilde{p}(w\vert x_0, \theta)
    &=
    p_{SDE}(g(w)\vert x_0, \theta)
    \left\vert
    \operatorname{det}
    \left(
    \nabla g(w)
    \right)
    \right\vert
    \nonumber
    \\
    &\propto
    \exp
    \left\{
    -
        \beta
        \sum_{i=0}^{N-1}
        \Big\Vert
        \frac{
        \hat{x}
        \left(t_{i+1}; w, x_0
        \right)-
        \hat{x}
        \left(t_{i}; w, x_0
        \right)
        }{\Delta t}
         - 
         f\left(
         \hat{x}
        \left(t_{i}; w, x_0
        \right), t_i; \theta
         \right)
        \Big\Vert^2
        \Delta t
    \right\}
    \left\vert
    \operatorname{det}
    \left(
    \nabla g(w)
    \right)
    \right\vert.
    \label{ptw}
\end{align}

Since we use the linear basis in \qref{repara_x} to parameterize $\hat{x}(t_{1:N};w, x_0)$, the transformation is linear, i.e., $x_{1:N} = A w+b(x_0)$. $A$ is a diagonal block matrix:
\begin{align*}
    A=
    \left[
    \begin{array}{c:c:c}
         \Phi_1&\cdots&\mathbf{0}_{N\times N}
         \\
         \vdots&\ddots&\vdots
         \\
         \mathbf{0}_{N\times N}&\cdots&\Phi_{d_x}
    \end{array}
    \right],
\end{align*}
where for $k=1, \cdots, d_x$ (recall $\psi_i$ is the dependent basis to enforce the initial condition)
\begin{align*}
    \Phi_k
    =
    \left[
    \begin{array}{ccc}
         \psi_{1}(t_1)-\frac{\psi_{1}(0)}{\psi_i(0)}\psi_i(t_1)
         &
         \cdots
         &
         \psi_{K}(t_1)-\frac{\psi_{K}(0)}{\psi_i(0)}\psi_i(t_1)
         \\
         &
         \ddots
         &
         \\
         \psi_{1}(t_N)-\frac{\psi_{1}(0)}{\psi_i(0)}\psi_i(t_N)
         &
         \cdots
         &
         \psi_{K}(t_N)-\frac{\psi_{K}(0)}{\psi_i(0)}\psi_i(t_N)
    \end{array}
    \right].
\end{align*}
The shift $b(x_0)$ is a constant vector depending on $x_0$.

It is easy to see $\nabla g(w; t_{1:N}, x_0)=A$.
Therefore
\begin{align*}
    \left\vert
\operatorname{det}
\left(
\nabla g(w; t_{1:N}, x_0)
\right)
\right\vert=
\text{constant}.
\end{align*}
So we can drop the determinant term in \qref{ptw} to derive:
\begin{align*}
    \tilde{p}(w\vert x_0, \theta)
    =
    \frac{
    \exp\left\{
    -
        \beta
        \sum_{i=0}^{N-1}
        \Big\Vert
        \frac{
        \hat{x}
        \left(t_{i+1}; w, x_0
        \right)-
        \hat{x}
        \left(t_{i}; w, x_0
        \right)
        }{\Delta t}
         - 
         f\left(
         \hat{x}
        \left(t_{i}; w, x_0
        \right), t_i; \theta
         \right)
        \Big\Vert^2
        \Delta t
    \right\}
    }{
    \tilde{Z}_{\beta}(x_0, \theta)
    },
\end{align*}
where
\begin{align*}
     \tilde{Z}_{\beta}(x_0, \theta)
     =
     \int_{\mathbb{R}^{d_w}} 
     dw \
     \exp\left\{
    -
        \beta
        \sum_{i=0}^{N-1}
        \Big\Vert
        \frac{
        \hat{x}
        \left(t_{i+1}; w, x_0
        \right)-
        \hat{x}
        \left(t_{i}; w, x_0
        \right)
        }{\Delta t}
         - 
         f\left(
         \hat{x}
        \left(t_{i}; w, x_0
        \right), t_i; \theta
         \right)
        \Big\Vert^2
        \Delta t
    \right\}.
\end{align*}
In our proposed physics-informed conditional prior in \qref{conditionalprior},
we apply the approximation
\begin{align*}
    \int_{0}^T dt\ \lVert
     \dot{x}(t)-f(x, t;\theta) \lVert^2
    =
    \sum_{i=0}^{N-1}
    \Big\Vert
    \frac{
    \hat{x}
    \left(t_{i+1}; w, x_0
    \right)-
    \hat{x}
    \left(t_{i}; w, x_0
    \right)
    }{\Delta t}
     - 
     f\left(
     \hat{x}
    \left(t_{i}; w, x_0
    \right), t_i; \theta
     \right)
    \Big\Vert^2
    \Delta t
    +
    O(\Delta t^2).
\end{align*}

The bijectivity of $g(w; t_{1:N}, x_0)$ can be achieved by choosing a more local basis, such as a radial basis.
Then $\Phi_k$ is a diagonally dominant matrix, so we can use the Gershgorin circle theorem \cite{golub2013matrix} to guarantee that $A$ is non-singular.

Following the same steps backwards, for any linear basis under the assumption of a bijective $g(w; t_{1:N})$, shows that the induced joint probability density of the state values at equidistant time points is independent of the choice of the basis.

Another important final remark is that one cannot simply make $\hat{x}(t;w,x_0)$ a nonlinear function of $w$, e.g., a neural network.
Doing so, will yield results that depend on the particular choice of the parameterization (both for the state values at equidistant time points and the parameters).
For non-linear parameterizations, it is hard to argue that the map between parameters to state values is bijective.
Furthermore, even if this map is bijective, one would have to multiply with the correct Jacobian determinant.

\section{
Parameterizing the truncated Fourier series using the initial point
}
\label{appendix:fourier}

We denote the $d_x \times (2N)$ matrix of the remaining coefficients by:
$$
 W = [w_1 \ w_2 \ \dots\  w_{2N}],
$$
and the corresponding basis functions by:
$$
 \psi_{-0}
(t)
=
 \begin{bmatrix}
    \cos{\left(\frac{2\pi  t}{T_{\text{max}}}\right)}&
    \sin{\left(\frac{2\pi  t}{T_{\text{max}}}\right)}&
    \cdots&
    \cos{\left(\frac{2\pi N t}{T_{\text{max}}}\right)}&
    \sin{\left(\frac{2\pi N t}{T_{\text{max}}}\right)}
 \end{bmatrix}^{T},
$$
where we remove the constant basis
$\psi_0 (t) = 1$.
Next, we denote the Fourier coefficients for the cosine bases in $\psi(t)$ by the $d_x\times N$ matrix:
\begin{align*}
    \hat{W} = [w_1\ w_3\ \dots\ w_{2N-1}]\in \mathbb{R}^{d_x \times N}.
\end{align*}
After conditioning $\Tilde{W}\psi(t)$ on the initial point $x(0) = x_0$ at $t=0$,  we have the equality constraints
\begin{align}
    w_1 =x_0-W\psi_{-0}(0) \nonumber
    =
    x_0-\hat{W}\boldsymbol{1},
    \nonumber
\end{align}
where $\boldsymbol{1}$ denotes a $K$-dimensional vector with all entries equal to $1$.

If we vectorize the Fourier coefficient matrix into a $d_w=d_x\times 2N$- dimensional vector $w = \operatorname{Vec}(W)$,
the parameterized function $\hat{x}(t; w, x_0)$ has the form of
\begin{align}
    \hat{x}(t;w, x_0) =
    x_0-\hat{W}\boldsymbol{1}
    +
    W \psi_{-0}(t).
    \label{fourier_gtW}
\end{align}
We see that the dependent basis term is simplified to
$
x_0-\hat{W}\boldsymbol{1}
$
when using the truncated Fourier series since the sine bases are 0 at $t=0$.

\section{The Gradient of ELBO}
\label{appendix_grad_ELBO}
We use the identity 
$
\nabla_{x}f(x) = 
f(x)\nabla_{x}\log\left(f(x)\right) 
$ to compute the gradient of the partition function with respect to $\psi$.
We have:
\begin{align} 
    \nabla_{\psi}
    \E{
    \log{
    Z_{\beta}\left(g_{\psi}\left(\eta\right)\right)
    }
    }
    &=
    \E{
        \frac{\nabla_\psi 
        Z_{\beta}\left(g_{\psi}\left(\eta\right)\right)
        }{
        Z_{\beta}\left(g_{\psi}\left(\eta\right)\right)
        }
    } \nonumber
    \\
    &=
    \E{
        \frac{\nabla_\psi\int d{\tilde{w}}\exp\left\{-\beta H\left(\wpp|g_{\psi}\left(\eta\right)\right)\right\}}
        {Z_{\beta}\left(
        g_{\psi}\left(\eta\right)
        \right)}
    } \nonumber
    \\
    &=
    \E{
        \frac{\int d{\tilde{w}}
        \left\{
        -\beta \nabla_{\psi}H\left(\wpp|g_{\psi}\left(\eta\right)\right)
        \right\}
        \exp\left\{-\beta H\left(\wpp|g_{\psi}\left(\eta\right)\right)\right\}}
        {Z_{\beta}\left(
        g_{\psi}\left(\eta\right)
        \right)}
    } \nonumber
    \\
    &=
    -
    \E{
    \E
    {\beta \nabla_{\psi}H\left(\wpp|g_{\psi}\left(\eta\right)\right)
    \middle | \eta
    }
    } \nonumber
    \\
    &=
    -
    \E{
    \E
    {\nabla_\psi h_\beta(\tilde{w},t|g_\psi(\eta))
    \middle| \eta
    }
    }.
    \nonumber
\end{align}

\section{Auxiliary guide as an approximate prior sampler}
\label{appendix_auxiliary_guide_appro_prior}
Starting from the partition function of the physics-informed conditional prior, we have
\begin{align}
    \log Z_{\beta}(w, \lambda)
    &=
    \bE_{w\sim q_{\phi}}
    \log Z_{\beta}(w, \lambda) \nonumber
    \\
    &=
    \bE_{w\sim q_{\phi}}
    \log \left\{
        \frac{
            \exp\left\{
                -\beta H\left(w|\lambda\right)
            \right\}
        }{
           p_{\beta}
           \left(
           w\vert \lambda
           \right)
        }
    \right\} \nonumber
    \\
    &=
    \bE_{w\sim q_{\phi}}
    \log \left\{
        \frac{
            \exp\left\{
                -\beta 
                H\left(w\vert \lambda\right)
            \right\}
        }{
            q_{\phi}(w)
        }
    \right\}
    +
    D_{KL}\left(
            q_{\phi}(w) \big \Vert 
            p_{\beta}
           \left(
           w\vert \lambda
           \right)
        \right). \nonumber
\end{align}
One can find that the first term in the last equality is the non-reparameterized version of \qref{vari_prior}.

\section{The maxmini variational inference}
\label{appendix_maxmini_vi}
In this appendix, we disclose that our Algorithm \ref{alg:vi_post} is a special and parsimonious case of a maxmini optimization problem.
We will only work on the non-reparameterized ELBO, and the extension to the reparameterized case is trivial.

Recall our ELBO has the form of
\begin{align}
    \operatorname{ELBO}\left(\phi, \psi\vert y\right)
    =
    \bE_{
    w\sim q_{\phi}, \lambda\sim q_{\psi}
    }
    \left[
    \log
    \left\{
        \frac{
            p\left(y|w, \lambda\right)
            \exp\big\{
                -\beta H\left(w|\lambda\right)
            \big\}
            p\left(\lambda\right)
        }
        {
           q_{\phi}(w) q_{\psi}(\lambda)
        }
    \right\}
    \right]
    -
    \bE_{\lambda \sim q_{\psi}}
    \left[
        \log 
        Z_{\beta}(\lambda)
    \right]. \label{ELBOwithZ_app}
\end{align}
 We introduce an amortized auxiliary guide 
$
q_{\Tilde{\phi}}(\Tilde{w}|\lambda)
$
,  and apply Jensen's inequality to lower bound the second term on the right-hand side of \qref{ELBOwithZ_app}:
\begin{align}
    \bE_{\lambda\sim q_{\psi}}
    \left[
        \log 
        Z_{\beta}(\lambda)
    \right]
    &=
    \bE_{\lambda\sim q_{\psi}}
    \left[
        \log 
        \int d\wpp \
        \exp\left\{
                -\beta H(\wpp|\lambda)
            \right\}
    \right] \nonumber
    \\
    &=
    \bE_{\lambda\sim q_{\psi}}
    \left[
        \log 
        \left\{
        \bE_{\wpp \sim q_{\Tilde{\phi}}(\wpp|\lambda)}
        \left[
            \frac{
            \exp\left\{
                -\beta H(\wpp|\lambda)
                \right\}
                }
                {q_{\Tilde{\phi}}\left(\Tilde{w}|\lambda\right)}          
        \right]
        \right\}
    \right] \nonumber
    \\
    &\geq
    \bE_{
    \lambda\sim q_{\psi},
    \wpp \sim q_{\Tilde{\phi}}(\wpp|\lambda)
    }
    \left[
    \log 
    \left\{
    \frac{
        \exp\left\{
        -\beta H(\wpp|\lambda)
        \right\}
        }
        {q_{\Tilde{\phi}}(\Tilde{w}|\lambda)} 
    \right\}
    \right].
    \label{lowerbound_Z}
\end{align}

Plug the inequality (\ref{lowerbound_Z}) into the ELBO in \qref{ELBOwithZ_app}, we derive an upper bound of the ELBO (UELBO)
\begin{align}
    \operatorname{UELBO}(\phi, \psi, \Tilde{\phi}|y)
    &=
    \bE_{w\sim q_{\phi}, \lambda\sim q_{\psi}}
    \left[
    \log
    \left\{
        \frac{
            p(y|w, \lambda)
            \exp\left\{
                -\beta H(w|\lambda)
            \right\}
            p(\lambda)
        }
        {
            q_{\phi}(w)q_{\psi}(\lambda)
        }
    \right\}
    \right] 
    -
    \bE_{
    \lambda\sim q_{\psi},
    \wpp \sim q_{\Tilde{\phi}}(\wpp|\lambda)
    }
    \left[
    \log 
    \left\{
    \frac{
        \exp\left\{
        -\beta H(\wpp|\lambda)
        \right\}
        }
        {q_{\Tilde{\phi}}(\Tilde{w}|\lambda)} 
    \right\}
    \right]
    \label{UELBO}.
\end{align}
So we have the inequalities
\begin{align}
    \log p(y)
    \geq
    \operatorname{ELBO}(\phi, \psi\vert y)
    \leq
    \operatorname{UELBO}(\phi, \psi, \Tilde{\phi}\vert y).
    \label{inequalites}
\end{align}
The inconsistent directions of the inequalities are discouraging at first moment. 
However, instead of maximizing the ELBO as in the standard VI, we can use the inequalities (\ref{inequalites}) to formulate a maxmini variational problem:
\begin{align}
    \max_{\phi, \psi}
    \
    \min_{\Tilde{\phi}}
    \
    \operatorname{UELBO}(\phi, \psi, \Tilde{\phi}\vert y).
    \label{maxmini_q}
\end{align}
This formulation aims to find the best amortized auxiliary guide $q_{\Tilde{\phi}}(\Tilde{w}|\lambda)$ to approximate the expected logarithmic partition function 
$
\bE_{\lambda\sim q_{\psi}}
\left[
    \log 
    Z_{\beta}(\lambda)
\right]
$
by minimizing UELBO,
and the best guide $q_{\phi}(w)$ and $q_{\psi}(\lambda)$ to approximate the posterior distribution $p(w, \lambda|y)$ by maximizing UELBO.

This general formulation is however computationally expensive since it requires solving an amortized auxiliary variational inference. 
More specifically, we should learn a family of guides indexed by $\lambda$.
The auxiliary guide $q_{\Tilde{\phi}}(\Tilde{w}|\lambda)$ can be parametrized by a neural network, and one can train this by incorporating an inner loop as in Algorithm \ref{alg:vi_post}.
However, in Algorithm \ref{alg:vi_post}, we don't train a costly amortized guide and instead dynamically update the auxiliary guide parameter $\Tilde{\phi}$ on the fly whenever we draw a new sample $\lambda_1 = g_{\psi}\left(\eta_1\right)$.

\section{
Training data details
}

Table \ref{tab:train} summarizes the training data details.
\begin{table}[!h]
\centering
\caption{Training data details.}
\label{tab:train}
\begin{tabular}{ccccc}
\toprule[0.5mm]
& Time duration 
& Number of data  
& Initial state 
& \begin{tabular}[c]{@{}c@{}}Measurement standard \\ deviation (relative)\end{tabular}
\\
\midrule[0.25mm]
Examples: 4.1.1 and 4.2.1 
& 
50s
& 
101 
& (10, 10, 10)  
& 5\%                                        \\
Examples: 4.1.2 and 4.2.2 
& 
(10, 20, $\cdots$, 100)s 
& 
(21, 41, $\cdots$, 201) 
& 
(10, 10, 10)  
& 5\%                                        \\
Example: 4.1.3         
& 100s 
& 201 
& (10, 10, 10) 
& 5\%                                        \\
Example: 4.3
& 50s 
& 501 
& (1, 0) 
& 5\%     
\\
\bottomrule[0.5mm]
\end{tabular}
\end{table}